\documentclass[twocolumn,numberedappendix,twocolappendix]{openjournal}

\usepackage{xcolor}
\usepackage{textgreek}
\usepackage[utf8]{inputenc}
\usepackage[english]{babel}
\usepackage{amsmath}
\usepackage{bm}
\usepackage{physics}
\usepackage{siunitx}
\usepackage{afterpage}
\usepackage{booktabs}
\usepackage{multirow}

\definecolor{linkcolor}{rgb}{0.0,0.3,0.5}
\usepackage{hyperref}
\hypersetup{
    unicode,
    colorlinks=true,
    linkcolor=linkcolor,
    citecolor=linkcolor,
    filecolor=linkcolor,
    urlcolor=linkcolor,
}

\usepackage{cleveref}
\crefname{table}{Table}{Tables}
\crefname{figure}{Fig.}{Figs.}
\crefname{equation}{Eq.}{Eqs.}
\crefname{section}{Section}{Sections}
\crefname{subsection}{Section}{Sections}
\crefname{subsubsection}{Section}{Sections}
\makeatletter
\@ifundefined{cref@label@type@}{\crefname{}{Section}{Sections}}{}
\makeatother
\usepackage{color,colortbl}
\usepackage{tensind}
\tensordelimiter{?}
\DeclareGraphicsExtensions{.bmp,.png,.jpg,.pdf}
\usepackage{verbatim}
\usepackage[normalem]{ulem}
\usepackage{orcidlink}
\usepackage{soul}
\usepackage{acro}
\usepackage{xspace}

\definecolor{rvsdone}{HTML}{1F6FD0}

\definecolor{brightorange}{HTML}{FF7A00}

\newcommand{\dsorig}{\texttt{original-all}\xspace}
\newcommand{\dsall}{\texttt{updated-all}\xspace}
\newcommand{\dsmcp}{\texttt{updated-MCP}\xspace}
\newcommand{\dsours}{\texttt{updated-ours}\xspace}
\newcommand{\Msun}{M_\odot}
\newcommand{\DA}{D_\mathrm{A}}

\newcommand{\vrec}{V_\mathrm{rec}}
\newcommand{\Vext}{\bm{V}_{\rm ext}}
\newcommand{\MBH}{M_\mathrm{BH}}
\newcommand{\Cv}{C_v}
\newcommand{\Ca}{C_a}

\newcommand{\kmsec}{\ensuremath{\mathrm{km}\,\mathrm{s}^{-1}}}
\newcommand{\Mpc}{\ensuremath{\mathrm{Mpc}}}
\newcommand{\kmsecMpc}{\ensuremath{\kmsec\,\Mpc^{-1}}}
\newcommand{\kmsecyr}{\ensuremath{\mathrm{km}\,\mathrm{s}^{-1}\,\mathrm{yr}^{-1}}}
\newcommand{\degunit}{\ensuremath{\mathrm{deg}}}
\newcommand{\degmas}{\ensuremath{\mathrm{deg}\,\mathrm{mas}^{-1}}}
\newcommand{\degmassq}{\ensuremath{\mathrm{deg}\,\mathrm{mas}^{-2}}}
\newcommand{\muas}{\ensuremath{\mu\mathrm{as}}}
\newcommand{\TWOMPP}{2M\texttt{++}}
\newcommand{\Manticore}{\texttt{Manticore-Local}}

\defcitealias{Pesce2020}{P20}
\defcitealias{Carrick2015}{C15}

\graphicspath{ {./figs/} }

\begin{document}

\title{A reanalysis of the megamaser Hubble constant: from spot catalogues to peculiar velocities}

\author{Richard Stiskalek\orcidlink{0000-0002-0986-314X}}
\email{richard.stiskalek@physics.ox.ac.uk}
\affiliation{Astrophysics, University of Oxford, Denys Wilkinson Building, Keble Road, Oxford, OX1 3RH, UK}
\author{Harry Desmond\orcidlink{0000-0003-0685-9791}}
\affiliation{Institute of Cosmology and Gravitation, University of Portsmouth, Dennis Sciama Building, Burnaby Road, Portsmouth, PO1 3FX, UK}

\begin{abstract}
The Hubble tension motivates careful scrutiny of the redshift-independent distances underpinning the local distance ladder.
Water megamasers afford a purely geometric distance from orbital dynamics in an accretion disc.
For five of the six Megamaser Cosmology Project (MCP) galaxies, we present a Bayesian forward model that infers the angular-diameter distance and warped Keplerian disc parameters jointly from the very long baseline interferometry quasi-observables: maser spot positions, velocities, and accelerations.
We then constrain $H_0$ from these distances and the host redshifts, including a physical distance prior, modelling the galaxy sample selection
and testing a linear and a non-linear peculiar-velocity reconstruction.
Accounting for peculiar velocities with the non-linear \Manticore\ reconstruction, we infer $H_0 = 71.6 \pm 3.7~\kmsecMpc$, where the uncertainty is statistical.
Relative to ignoring peculiar velocities, the linear and non-linear reconstructions lower $H_0$ by $0.8$--$1.4$ and $1.6$--$2.1~\kmsecMpc$, respectively, whereas changing the selection variable alters it by
less than $1~\kmsecMpc$.
We analyse the sixth galaxy, NGC~4258, separately. Modelling its disc as eccentric and quadratically warped, we infer an angular-diameter distance of $7.558 \pm 0.073\,\mathrm{(stat.)}~\Mpc$, within $0.13\,\sigma$ of the published maser distance.
\end{abstract}


\maketitle

\section{Introduction}\label{sec:intro}

The Hubble constant $H_0$ describes the present-day expansion rate of the Universe and anchors the extragalactic distance scale.
Inferences from the cosmic microwave background (CMB) under flat $\Lambda$CDM yield $H_0 = 67.4 \pm 0.5~\kmsecMpc$~\citep{Planck2020cosmo}, while local distance-ladder measurements of the SH0ES programme, using Cepheid-calibrated Type~Ia supernovae, give $H_0 = 73.2 \pm 0.9~\kmsecMpc$~\citep[][see also~\citealt{Riess2022}]{Breuval2024}.
An alternative calibration of Type~Ia supernovae by the Carnegie--Chicago Hubble Program gives a lower $H_0 = 70.4 \pm 1.9~\kmsecMpc$~\citep{Freedman2025}, intermediate between the two.
The discrepancy between the \textit{Planck} and SH0ES $H_0$ values in particular has been dubbed the ``Hubble tension'', and has prompted debate over whether it signals new physics beyond $\Lambda$CDM or unrecognised systematic or statistical errors in one or both inference pipelines~\citep{Verde2019,DiValentino2021,Freedman2021,CosmoVerse2025}.
Understanding the tension further requires independent approaches with rigorous statistical inference methodologies.

Water megamasers in the accretion discs of supermassive black holes in active galactic nuclei (AGN) provide such a method~\citep{Lo2005}.
The 22~GHz maser emission, collisionally pumped in dense molecular gas irradiated by the central X-ray source~\citep{Neufeld1994,Neufeld1995,Collison1995,Maloney2002}, traces a thin, sub-parsec Keplerian disc whose geometry and kinematics are measured by very long baseline interferometry (VLBI) imaging and multi-epoch spectral monitoring~\citep{Miyoshi1995,Herrnstein1999}.
Such discs have been resolved in the archetypal NGC~4258 and in a handful of other AGN~\citep{Greenhill1996,Greenhill2003,Kondratko2005,Kondratko2008}.
Maser spot velocities constrain the ratio of black hole mass to angular-diameter distance, $\MBH / \DA$, while centripetal accelerations constrain $\MBH / \DA^2$.
Together, these break the mass--distance degeneracy and yield a purely geometric distance estimate~\citep{Reid2014, Reid2019}.

The Megamaser Cosmology Project~\citep[MCP;][]{Braatz2007,Reid2009} measured geometric distances to six megamaser-hosting galaxies over more than a decade of VLBI imaging and spectral monitoring~\citep[e.g.][]{Kuo2011,Humphreys2013,Reid2013,Kuo2013,Kuo2015,Gao2016,Pesce2020b}.
In the culminating analysis,~\citet{Pesce2020} (hereafter~\citetalias{Pesce2020}) re-derive the distances to four of these galaxies with an updated sampling procedure, adopting the published values for CGCG~074-064~\citep{Pesce2020b} and NGC~4258~\citep{Reid2019}, and combine them to obtain $H_0 = 73.9 \pm 3.0~\kmsecMpc$, a $4.1$ per cent constraint consistent with local distance-ladder results and in ${\sim}2\,\sigma$ tension with the CMB.
\citet{Boruah2021} reanalyse the MCP distances with a reconstructed peculiar-velocity field~\citep[hereafter~\citetalias{Carrick2015}]{Carrick2015} from the \TWOMPP\ galaxy redshift survey~\citep{Lavaux2011}, marginalising over the line-of-sight peculiar velocity, and obtain $H_0 = 70.1 \pm 2.9~\kmsecMpc$.
Their reanalysis improves only the distance-to-$H_0$ step, taking the~\citetalias{Pesce2020} geometric-distance posteriors as fixed inputs.
Most recently,~\citet{vanderBoom2025} propose an alternative method for NGC~4258, tracking individual maser components across epochs, and suggest that high-cadence monitoring is essential for reliable maser distance estimates.

The~\citetalias{Pesce2020} analysis proceeds in two stages: the warped Keplerian disc model is first applied to the VLBI data for each galaxy independently to obtain a posterior on $\DA$, and the six distance posteriors are then combined with recession velocities to constrain $H_0$.
Stage~1 contains all the geometric information, and its posterior is non-trivial to explore: the per-spot azimuthal angles are bimodal and the mass--distance degeneracy forms a curved ridge.
In addition, neither this analysis nor the reanalysis of~\citet{Boruah2021} models the selection of the megamaser sample, which can bias distance-ladder inferences of $H_0$~\citep{Stiskalek2026a,Desmond2026}.

Another important input to stage~1 is which maser spots enter the disc model and what measurements those spots have.
For four of the five galaxies we analyse, the~\citetalias{Pesce2020} disc modelling uses spot tables that, following a series of internal quality checks, differ from and supersede the catalogues reported in the preceding MCP papers~\citep{Pesce2026}.\footnote{A preprint of the erratum is available at \url{https://github.com/dpesce/MCP-H0-erratum}.}
These tables differ in two ways.
The first is \emph{table composition}: which spots are present at all, with outlier rejection to remove non-physical emission regions and, for one galaxy, the addition of spots from an earlier epoch.
The second is \emph{table content}: what values the spots have, with updates to quantities such as the per-spot acceleration uncertainties and sky positions.
We therefore report every result for two baselines, both using the updated measurements: the~\dsmcp table---the input used by the~\citetalias{Pesce2020} disc modelling---and the~\dsours table, the same measurements with our own iterative clipping in place of the MCP outlier rejection, and a different error-floor model for NGC~5765b.

In this work, we present a Bayesian forward model for the MCP sample, built on the same warped Keplerian disc physics assumed throughout the MCP literature, that infers each galaxy's distance and disc parameters directly from the VLBI quasi-observables---maser spot sky positions, line-of-sight (LOS) velocities, and LOS accelerations---and constrains $H_0$ from the distances and host recession velocities.
The key methodological advances are:
\begin{enumerate}
    \item A global optimisation of each per-galaxy disc posterior that resolves the azimuthal bimodality, including numerical marginalisation of the per-spot orbital coordinates on non-uniform grids, reducing the per-galaxy mode search from ${\sim}2N_\mathrm{spots} + 14$ to $14$ parameters.
    \item A Metropolis-within-Gibbs sampler that updates the per-spot orbital coordinates in dedicated Gibbs blocks and jumps between their reflected azimuthal solutions during sampling.
    \item The use of physical uniform-in-volume priors for galaxy distances, rather than priors flat in distance or distance modulus which bias distances low.
    \item A first, intentionally simple model of the sample selection, absent from previous megamaser $H_0$ analyses, in which a soft threshold enters the standard selection-corrected hierarchical posterior.
    \item A quantification of the impact of the MCP reanalysis of the spot measurements, propagating four spot tables end to end through the same inference: our two baselines, which share the updated MCP measurements and differ in whether the outlying spots are removed by the MCP or by our own clipping and, for NGC~5765b, in the error-floor model, together with the superseded published catalogues and an uncut updated table, which together bound the effect of the updated values against that of the removed or added spots.
    \item A peculiar-velocity treatment that evaluates the reconstructed line-of-sight velocity at the sampled distance and marginalises over the constrained realisations of the local Universe, instead of assigning each host a single fixed velocity correction.
\end{enumerate}
Our implementation reproduces the \texttt{fit\_disk} likelihood of~\citet{Reid2013} to machine precision, so our disc model and likelihood are those of the MCP analyses, and any distance difference must originate in the input table, the priors, or the posterior convergence.
    On the~\dsmcp baseline our distances agree with their values to within $0.3\,\sigma$ for every galaxy, and on the \dsours baseline to within $1.0\,\sigma$.

The remainder of this paper is organised as follows.
Section~\ref{sec:data} describes the data.
Section~\ref{sec:method} introduces the forward model, including the warped Keplerian disc physics, likelihood, priors, $H_0$--distance connection, and sample selection treatment.
Section~\ref{sec:results} presents results from the MCP sample: the per-galaxy disc posteriors (Section~\ref{sec:results_disc}), the impact of the MCP reanalysis on the inferred distances (Section~\ref{sec:origin_shift}), the Hubble constant inference and our reproduction of the~\citetalias{Pesce2020} population model (Section~\ref{sec:results_h0}), and NGC~4258 (Section~\ref{sec:ngc4258}).
Section~\ref{sec:discussion} discusses open issues, and Section~\ref{sec:conclusion} concludes.
Throughout, $\mathcal{N}(\mu, \sigma)$ denotes a normal distribution with mean $\mu$ and standard deviation $\sigma$; $\pi(\cdot)$ a prior density; $\mathcal{L}$ a likelihood; and $h \equiv H_0/(100~\kmsecMpc)$ the reduced Hubble constant.
Observed quantities have the subscript ``obs'', and $\log$ denotes the base-10 logarithm.
A difference between two posteriors is quoted in units of the quadrature sum of their symmetrised posterior half-widths, where the symmetrised half-width is the mean of the upper and lower $68\%$ bounds.

\section{Data}\label{sec:data}

The MCP observed six megamaser-hosting galaxies, whose distances~\citetalias{Pesce2020} combine into the final $H_0$ constraint.
These galaxies span recession velocities from $V_\mathrm{sys} \approx 679$ to ${\sim}\num{10200}~\kmsec$, or distances from $D \approx 7.6$ to $150~\Mpc$.
Within our forward model, we work with the VLBI quasi-observables rather than derived distances, adopting the MCP-reported per-spot measurements directly without re-reducing the raw interferometric data.
Of the six~\citetalias{Pesce2020} galaxies, we exclude NGC~4258, the nearest in the sample and the prototype of the maser-disc class~\citep{Nakai1993,Watson1994,Greenhill1995,Maoz1995}, from the maser-only $H_0$ inference: at $D \approx 7.6~\Mpc$ its recession velocity ($V_\mathrm{sys} \approx 679~\kmsec$) is comparable to typical peculiar velocities and contains little information about $H_0$.
Its geometric maser distance nonetheless anchors the local distance ladder~\citep{Reid2019}, calibrating, for example, the Cepheid zero-point of SH0ES~\citep{Riess2022} and the zero-point of the tip of the red giant branch in the Carnegie--Chicago Hubble Program~\citep{Freedman2019}.
We therefore analyse it separately in Section~\ref{sec:ngc4258}, outside the joint $H_0$ posterior.
We use the data of~\citet{Reid2019}, obtained from M.~Reid (private communication), which contains $358$ spots---$187$ systemic, $32$ approaching, and $139$ receding---of which $151$ have a measured acceleration.
\Cref{tab:data_availability} summarises the data availability for each galaxy.
All five spot catalogues tabulate per-spot velocities in the optical convention, so no radio-to-optical correction is required.\footnote{The excluded NGC~4258 is the only MCP galaxy whose maser velocities are tabulated in the radio convention; we convert these with $V_\mathrm{opt} = V_\mathrm{radio}/(1 - V_\mathrm{radio}/c)$.}
The spot velocities are referred to the local standard of rest for UGC~3789, NGC~5765b, NGC~6264, NGC~6323, and NGC~4258, and to the barycentric frame for CGCG~074-064.
We convert only the galaxy recession velocity $V_\mathrm{sys}$ to the CMB frame~\citep{Planck_2020}.
We sample a per-galaxy systemic offset $\Delta V_\mathrm{sys}$ that accounts for any constant additive shift between velocity frames, leaving the inference insensitive to the choice of the local standard of rest or the barycentric frame.

\begin{table*}
    \centering
    \caption{Summary of the maser spot catalogues for the six MCP galaxies.
    The reference column gives the MCP VLBI and spectral-monitoring papers from which each spot catalogue is taken.
    Spots are classified by their LOS velocity into systemic features, and approaching (blue) and receding (red) high-velocity features.
    Each class entry gives the number of spots with a measured LOS acceleration over the total, $n_\mathrm{acc}/n_\mathrm{tot}$, and every spot additionally has a sky-plane position and an LOS velocity.
    $V_\mathrm{sys}$ is the CMB-frame recession velocity.
    The last two count columns give the number of spots in the catalogue as printed in the source paper (\dsorig) and the number retained in the updated table used by~\citetalias{Pesce2020} (\dsmcp); see Section~\ref{sec:datasets}. NGC~4258 has a single table.}
    \label{tab:data_availability}
    \begin{tabular*}{\textwidth}{@{\extracolsep{\fill}}llcccccc@{}}
        \toprule
        & & \multicolumn{3}{c}{Spots by class [$n_\mathrm{acc}/n_\mathrm{tot}$]} & & & \\
        \cmidrule(lr){3-5}
        Galaxy & Reference & Systemic & Blue & Red & \dsorig & \dsmcp & $V_\mathrm{sys}$ [$\kmsec$] \\
        \midrule
        NGC~4258 & M.~Reid (priv.\ comm.) & 95/187 & 12/32 & 44/139 & \multicolumn{2}{c}{358} & 679 \\
        UGC~3789 & \citet{Reid2013} & 42/42 & 26/68 & 12/46 & 156 & 153 & \num{3320} \\
        CGCG~074-064 & \citet{Pesce2020b} & 45/45 & 30/49 & 70/71 & 165 & 165 & \num{7172} \\
        NGC~6323 & \citet{Kuo2011,Kuo2015} & 11/11 & 4/21 & 14/36 & 68 & 87 & \num{7802} \\
        NGC~5765b & \citet{Gao2016} & 40/60 & 12/79 & 17/73 & 212 & 169 & \num{8526} \\
        NGC~6264 & \citet{Kuo2011,Kuo2013} & 11/11 & 12/32 & 14/23 & 66 & 61 & \num{10193} \\
        \bottomrule
    \end{tabular*}
\end{table*}

\begin{figure*}
    \centering
    \includegraphics[width=\textwidth]{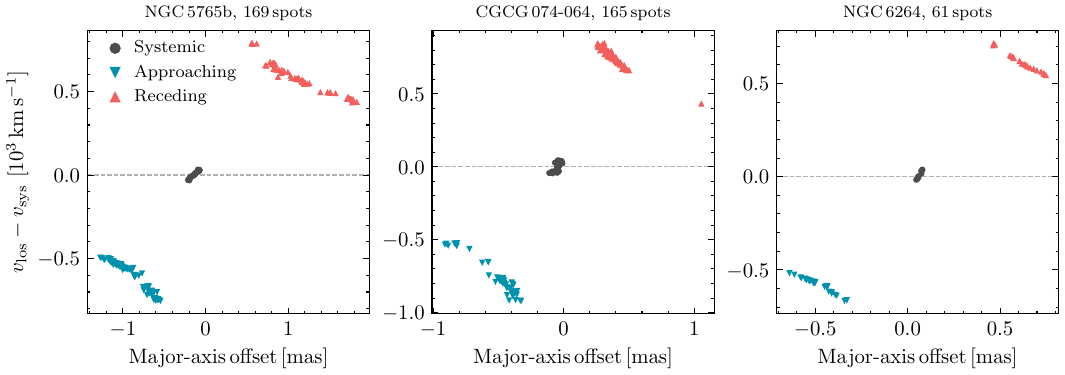}
    \caption{Position--velocity diagrams of the maser spots in NGC~5765b, CGCG~074-064, and NGC~6264, coloured by spectral class, for the~\dsmcp tables (Section~\ref{sec:datasets}).
    The vertical axis is the LOS velocity offset from the systemic velocity.
    For illustration, the horizontal axis is the maser sky position projected onto the disc major axis, which we take empirically as the direction of maximum variance of the spot positions.
    The systemic spots cluster near the origin, in front of the black hole, while the approaching and receding high-velocity spots populate the disc tangents at opposite ends of the major axis.}
    \label{fig:spot_classification}
\end{figure*}

The five maser spot catalogues differ in completeness but share a common structure: every spot has a sky-plane position and an LOS velocity, and a subset additionally has a measured LOS acceleration.
Every systemic spot has a measured acceleration except in NGC~5765b, where $20$ of the $60$ systemic spots lack it.
Accelerations are by contrast sparse among the fainter high-velocity features, reflecting the flux-limited, multi-epoch observations.
We classify each spot by clustering the LOS velocities into three groups: systemic features near $V_\mathrm{sys}$, where the disc midline crosses the line of sight in front of the black hole, and approaching (blue) and receding (red) high-velocity features Doppler-shifted by several hundred~$\kmsec$ to either side.
In~\cref{fig:spot_classification} we show this classification for NGC~5765b, CGCG~074-064, and NGC~6264, where the three populations separate cleanly.

The systemic spectrum of NGC~5765b divides into two velocity clumps, and~\citet{Gao2016} treat them separately when measuring the accelerations.
The ``clump~1'', between $\num{8260}$ and $\num{8290}~\kmsec$, has distinct and well-measured lines, whereas the ``clump~2'', between $\num{8290}$ and $\num{8340}~\kmsec$, is more blended.
The two clumps together contain the $40$ systemic spots with a measured acceleration, $17$ in clump~1 and $23$ in clump~2, the remaining $20$ systemic spots lacking acceleration measurements.
We adopt the~\citet{Gao2016} clump~2 velocity range and use a second set of error floors in the model for it.

The catalogues report no per-spot velocity uncertainties, so we assign a fixed measurement floor of $0.25~\kmsec$ and add it in quadrature to a sampled error floor $\sigma_v$ in~\cref{eq:L_vel}.
This sampled floor dominates the velocity error budget, with the fixed measurement floor serving only to keep the likelihood numerically stable.

\subsection{The maser spot tables}\label{sec:datasets}

The disc modelling of~\citetalias{Pesce2020} does not use the original MCP spot tables for four of the five galaxies analysed here, instead using tables produced by their own internal reanalysis and quality cuts.
These updated measurements come from the MCP's subsequent reanalysis of the same VLBI observations, and they supersede the values published in the preceding MCP papers, as the erratum to~\citetalias{Pesce2020} sets out~\citep{Pesce2026}.
The revision of the measured values is confined to two galaxies: the acceleration uncertainties of NGC~6264 and the sky positions of NGC~6323.
The differences from the original catalogues are the replacement of a fraction of the per-spot position and acceleration measurements with values from an internal MCP analysis, and the removal of outlying spots at approximately the $3\,\sigma$ level through a blinded, iterative procedure.

We therefore propagate four spot tables, two current and two diagnostic, end to end through the same disc model, priors, and sampler.
We name each table for the measurements it contains and the spots it retains, as \textit{measurements}-\textit{spots}.
The \dsorig tables are the catalogues as printed in the source papers cited in~\cref{tab:data_availability}, retained here to quantify the impact of the MCP reanalysis that supersedes them.
The \dsall tables have the updated MCP measurements but retain every spot, so they isolate the measurement update from the spot removal.\footnote{The isolation is not exact: the MCP-updated tables report no updated measurement for a spot it discards, so the $43$, $5$, and $3$ discarded spots of NGC~5765b, NGC~6264, and UGC~3789 retain their originally published values in~\dsall.}
The \dsmcp tables are the \dsall tables after the MCP cut, and are the input their disc modelling actually uses.
The \dsours tables are the \dsall tables after our own iterative outlier rejection (Section~\ref{sec:clipping}), so they have the updated measurements under a spot selection we define.
The last two therefore differ in the outlier rejection method used and, for NGC~5765b, in the error-floor model (Section~\ref{sec:disc_model}).
The~\dsmcp and \dsours tables are our two baselines, both built on the current measurements, whereas the \dsall and \dsorig tables decompose the shift from the published catalogues to the updated ones into a contribution from what measurements the spots have and a contribution from which spots are retained.
This decomposition is exact for every galaxy except NGC~6323, whose \dsall table also adds spots.

The differences are galaxy-specific:
\begin{enumerate}
    \item \textit{CGCG~074-064}. This is the only galaxy that remains unchanged between its presentation in~\citet{Pesce2020b} and the~\citetalias{Pesce2020} joint cosmological analysis.
    \item \textit{NGC~5765b}. \citetalias{Pesce2020} remove $43$ of the $212$ spots published by~\citet{Gao2016}, the largest cut in the sample.
    Of these, $20$ are the systemic features with placeholder accelerations, and $23$ are removed by the MCP iterative outlier rejection.
    \citetalias{Pesce2020} clip clump~2 down to $16$ spots from $28$, compared to $14$ from $32$ for clump~1, but in clump~2 the placeholder rows account for only $5$ of the $12$ spots removed, whereas in clump~1 they account for $15$ of the $18$.
    So while for clump~1 the MCP cut mainly removes spots with missing accelerations, in the more blended clump~2 it is predominantly their outlier rejection.
    \item \textit{UGC~3789}. \citetalias{Pesce2020} remove $3$ of the $156$ spots published by~\citet{Reid2013}.
    \item \textit{NGC~6264}. \citetalias{Pesce2020} remove $5$ of the $66$ spots published by~\citet{Kuo2013} and replace the acceleration uncertainties of $33$ spots with values from an internal MCP acceleration analysis, inflating them by a median factor of $3$.
    \item \textit{NGC~6323}. This is the one galaxy whose~\dsmcp table is larger than its \dsorig one: the $68$-spot table of~\citet{Kuo2015} is augmented with $19$ first-epoch spots from~\citet{Kuo2011}, and no spot is removed thereafter.
    The sky positions and positional uncertainties of $67$ of the original $68$ spots are also updated, with the sky positions shifting by up to $12$ and $38~\muas$ in the two sky coordinates.
\end{enumerate}

Our own clipping (Section~\ref{sec:clipping}) removes a comparable number of spots to the MCP cut, but a largely different set of them.
Of the $686$ spots in the five \dsall tables, we remove $13$, $6$, $8$, $1$, and $4$ from CGCG~074-064, NGC~5765b, UGC~3789, NGC~6264, and NGC~6323, respectively.
The MCP cut removes $51$ spots in total, of which $20$ are the NGC~5765b placeholder rows described above, so it clips $31$ spots, compared to $32$ by our own method, of which only $8$ are common to both.
As we shall show, this difference in clipping has a negligible effect on the inferred distances.

\section{Methodology}\label{sec:method}

We present the method in three parts.
In Section~\ref{sec:disc_model} we review the standard maser-disc analysis: conditioned on the galaxy's angular-diameter distance $\DA$, black hole mass $\MBH$, and disc geometry, a warped Keplerian disc forward model predicts the VLBI quasi-observables---spot sky positions, LOS velocities, and LOS accelerations---with no reference to $H_0$.
In Section~\ref{sec:sampling} we describe how we sample this posterior, whose bimodal per-spot azimuthal angles and warped disc geometry make locating the global optimum the central difficulty: a differential-evolution search first initialises the chains at the global posterior maximum, and a Metropolis-within-Gibbs scheme then alternates between the global disc parameters and the per-spot orbital coordinates, jumping between their reflected azimuthal solutions.
Finally, in Section~\ref{sec:joint_prob} we combine the five galaxies into a constraint on $H_0$: each galaxy's distance posterior is sampled jointly with the shared $H_0$ and peculiar-velocity scatter $\sigma_\mathrm{pec}$, and the inferred distances are compared to the host recession velocities.

\subsection{Warped Keplerian disc model}\label{sec:disc_model}

We first describe the model for a single galaxy with $N$ maser spots.
The baseline disc model parameters $\boldsymbol{\theta}$ comprise the angular-diameter distance $\DA$, the black hole mass $\MBH$, the disc geometry (the inclination $i_0$ and position angle $\Omega_0$ with their linear warp rates $\mathrm{d}i/\mathrm{d}r|_{r_\mathrm{ref}}$, $\mathrm{d}\Omega/\mathrm{d}r|_{r_\mathrm{ref}}$), the black hole sky position $(x_0, y_0)$, a systemic-velocity offset $\Delta V_\mathrm{sys}$, and the error-floor parameters for the spot positions ($\sigma_x$ and $\sigma_y$), velocities ($\sigma_{v,\mathrm{sys}}$ and $\sigma_{v,\mathrm{hv}}$, the non-thermal line broadening of systemic and high-velocity features, ${\sim}2$--$3~\kmsec$), and accelerations ($\sigma_a$).
For NGC~5765b, $\boldsymbol{\theta}$ also contains a separate set of the four error floors applied to the clump~2 systemic spots.
We include this second set of floors on the \dsorig, \dsall, and \dsours tables, but not on the~\dsmcp table, where the MCP cut has already removed $12$ of the $28$ clump~2 spots.

Two optional blocks extend the geometry: the quadratic warp rates $\mathrm{d}^2i/\mathrm{d}r^2|_{r_\mathrm{ref}}$ and $\mathrm{d}^2\Omega/\mathrm{d}r^2|_{r_\mathrm{ref}}$, and the orbital eccentricity $e$ with its argument of periapsis $\omega_0$ and warp rate $\mathrm{d}\omega/\mathrm{d}r|_{r_\mathrm{ref}^{\omega}}$.
The comoving distance follows from the angular-diameter distance as $D = \DA\,(1 + z_\mathrm{cosmo})$, with $z_\mathrm{cosmo}$ the cosmological redshift.
The disc likelihood depends on the distance only through $\DA$, so the per-galaxy inference makes no assumption about $H_0$ or the background cosmology.
A fiducial flat $\Lambda$CDM cosmology ($H_0 = 73~\kmsecMpc$, $\Omega_{\rm m} = 0.315$) enters only to set the non-informative bounds of the uniform $\DA$ prior (Section~\ref{sec:sampling}).
The Hubble constant is inferred entirely in stage~2 (Section~\ref{sec:joint_prob}).
The per-spot orbital coordinates---the radii $r_i$ and azimuthal angles $\phi_i$---are an intrinsic part of the disc model: they are marginalised out explicitly in the global mode search and drawn jointly with the global disc parameters during posterior sampling.
\Cref{tab:priors} lists all model parameters with their priors.
In place of $\MBH$ we sample the mass parameter
\begin{equation}\label{eq:eta}
    \eta \equiv \log \frac{\MBH/\Msun}{\DA/\Mpc},
\end{equation}
the distance-decorrelated combination of black hole mass and distance that the velocities constrain through~\cref{eq:vkep}.\footnote{We sample $\eta$ under a uniform prior.
    Because $\log(\MBH/\Msun) = \eta + \log(\DA/\Mpc)$ differs from $\eta$ only by the additive distance term, the change of variables has a unit Jacobian, $\partial\eta/\partial\log\MBH = 1$, so the flat prior on $\eta$ is equivalently a flat prior on $\log\MBH$ (\cref{tab:priors}).}
Sampling $\eta$ rather than $\MBH$ is motivated by the mass--distance degeneracy: the velocities constrain only the ratio $\MBH/\DA$, while the accelerations, which constrain $\MBH/\DA^2$, break it only partially, so in the original variables $\MBH$ and $\DA$ lie along a strongly correlated ridge.
This reparametrisation facilitates sampling.

\begin{table*}
    \centering
    \caption{Model parameters, descriptions, and prior distributions for each maser disc galaxy, where stage~1 is the per-galaxy disc inference and stage~2 the joint $H_0$ inference.
        $\mathcal{U}(a,b)$ denotes a uniform distribution, $\mathcal{N}(\mu,\sigma)$ a normal distribution (truncated to the listed support where one is given), and $\mathcal{S}(a,b)$ a sine prior $\propto \sin i$ on $[a,b]$.
        The optional blocks are sampled only in the corresponding quadratic-warp and eccentric model variants and are disabled in the baseline analysis.
        For clump~2 of NGC~5765b we model a second set of error floors.
        Deterministic rows are exact functions of the sampled parameters or fixed values, and square brackets give the induced prior.}
    \label{tab:priors}
    \begin{tabular}{lll}
        \toprule
        Parameter                                            & Description                                                      & Prior / transformation                                                                               \\
        \midrule
        \multicolumn{3}{l}{\textit{Disc parameters --- sampled}}                                                                                                                                                                       \\
        $\DA$                                                & Angular-diameter distance                                        & $\mathcal{U}(\DA^{\min}, \DA^{\max})$ at stage~1 (uniform-in-volume $D^2$ at stage~2)
        \\
        $\eta$                                               & Mass--distance parameter $\log[(\MBH/\Msun)/(\DA/\Mpc)]$ & $\log(\MBH/\Msun) \sim \mathcal{U}(4, 10)$                                                           \\
        $x_0, y_0$                                           & BH sky position                                                  & $\mathcal{N}(0, 500)~\muas$ on $[-750, 750]~\muas$                                                   \\
        $i_0$                                                & Disc inclination                                                 & $\mathcal{S}(65, 115)~\degunit$                                                                          \\
        $\Omega_0$                                           & Position angle                                                   & $\mathcal{U}(0, 360)~\degunit$                                                                           \\
        $\mathrm{d}i/\mathrm{d}r|_{r_\mathrm{ref}}$          & Inclination warp rate                                            & $\mathcal{U}(-30, 30)~\degmas$                                                                      \\
        $\mathrm{d}\Omega/\mathrm{d}r|_{r_\mathrm{ref}}$     & PA warp rate                                                     & $\mathcal{U}(-30, 30)~\degmas$                                                                      \\
        $\Delta V_\mathrm{sys}$                              & Systemic velocity offset                                         & $\mathcal{N}(0, 300)~\kmsec$                                                                         \\
        $\sigma_{x}, \sigma_{y}$                             & Position error floor                                             & $\mathcal{N}(10, 5)~\muas$ on $[0.5, 100]~\muas$                                                     \\
        $\sigma_{v,\mathrm{sys}}$                            & Systemic velocity error floor                                    & $\mathcal{N}(2, 1)~\kmsec$ on $[0.5, 100]~\kmsec$                                              \\
        $\sigma_{v,\mathrm{hv}}$                             & High-velocity error floor                                        & $\mathcal{N}(2, 1)~\kmsec$ on $[0.5, 100]~\kmsec$                                              \\
        $\sigma_{a}$                                         & Acceleration error floor                                         & $\mathcal{N}(0.3, 0.15)~\kmsecyr$ on $[0.02, 0.75]~\kmsecyr$                                   \\
        \midrule
        \multicolumn{3}{l}{\textit{Disc parameters --- sampled (optional)}}                                                                                                                                                            \\
        $\mathrm{d}^2i/\mathrm{d}r^2|_{r_\mathrm{ref}}$      & Inclination quadratic warp                                       & $\mathcal{N}(0, 90)~\degmassq$                                                                    \\
        $\mathrm{d}^2\Omega/\mathrm{d}r^2|_{r_\mathrm{ref}}$ & PA quadratic warp                                                & $\mathcal{N}(0, 90)~\degmassq$                                                                    \\
        $e_x, e_y$                                           & Eccentricity components                                          & $\mathcal{N}(0, 0.025)$                                                                              \\
        $\mathrm{d}\omega/\mathrm{d}r|_{r_\mathrm{ref}^{\omega}}$ & Periapsis warp rate                              & $\mathcal{U}(-360, 360)~\degmas$ for NGC~4258; fixed at $0$ otherwise                              \\
        \midrule
        \multicolumn{3}{l}{\textit{Deterministic}}                                                                                                                                                                                     \\
        $D$                                                  & Comoving distance                                                & $\DA\,(1+z_\mathrm{cosmo})$                                                                          \\
        $\log(\MBH/\Msun)$                                   & Black hole mass                                                  & $\eta + \log(\DA/\Mpc)$                                                                      \\
        $e$                                                  & Orbital eccentricity                                             & $(e_x^2 + e_y^2)^{1/2}~[\mathrm{Rayleigh}(0.025)]$                                                   \\
        $\omega_0$                                           & Argument of periapsis                                            & $\mathrm{atan2}(e_y, e_x)~[\mathcal{U}(0, 2\pi)]$                                                    \\
        \midrule
        \multicolumn{3}{l}{\textit{Per-spot (sampled)}}                                                                                                                                                                                \\
        $r_i$                                                & Orbital angular radius                                           & Flat, $r_i > 0$ (improper)                                                                           \\
        $\phi_i$                                             & Azimuthal angle                                                  & $\mathcal{U}(0, 2\pi)$ systemic, $\mathcal{U}(0,\pi)$/$\mathcal{U}(\pi,2\pi)$ red/blue high-velocity \\
        \bottomrule
    \end{tabular}
\end{table*}

The disc model maps orbital coordinates in the disc frame to quasi-observable positions, velocities, and accelerations in the sky frame.
The transformation between these frames is specified by two angles, corresponding to the direction of the disc normal vector.
These are the inclination $i$ (the angle between the disc normal and the LOS, with $i = 90^\circ$ corresponding to an edge-on disc) and the position angle $\Omega$ (the orientation of the disc major axis on the sky, measured from north through east to the receding side).
Real maser discs are not flat: tidal torques and Lense--Thirring precession~\citep{Bardeen1975,Caproni2006} cause $i$ and $\Omega$ to vary with orbital radius, as seen in the resolved warp of NGC~4258~\citep{Maloney1996,Martin2008}.
Following~\citet{Herrnstein2005} and later megamaser-disc analyses~\citep{Kuo2011, Reid2013, Humphreys2013, Gao2016, Pesce2020b}, we adopt a Taylor expansion to second order about a fixed pivot radius $r_\mathrm{ref}$, set to approximately the median spot radius to decorrelate the warp rates from their zero-points:
\begin{equation}\label{eq:warp_i}
    i(r) = i_0 + \left.\frac{\mathrm{d}i}{\mathrm{d}r}\right|_{r_\mathrm{ref}}\,(r - r_\mathrm{ref}) + \left.\frac{\mathrm{d}^2 i}{\mathrm{d}r^2}\right|_{r_\mathrm{ref}}\,(r - r_\mathrm{ref})^2,
\end{equation}
and
\begin{equation}\label{eq:warp_Omega}
    \Omega(r) = \Omega_0 + \left.\frac{\mathrm{d}\Omega}{\mathrm{d}r}\right|_{r_\mathrm{ref}}\,(r - r_\mathrm{ref}) + \left.\frac{\mathrm{d}^2 \Omega}{\mathrm{d}r^2}\right|_{r_\mathrm{ref}}\,(r - r_\mathrm{ref})^2,
\end{equation}
where $r$ is the angular orbital radius, the subscript denotes evaluation at the pivot radius $r_\mathrm{ref}$, the linear warp rates are in $\degmas$, and the quadratic coefficients in $\degmassq$.

Each maser spot lies at disc-frame coordinates $(r, \phi)$, where $\phi$ is the azimuthal angle measured from the near side of the disc midline ($\phi = 0$ corresponds to the point closest to the observer along the LOS).
In this convention the systemic features lie near $\phi = 0$ and $\phi = \pi$, where the orbital motion is transverse to the LOS, whereas the high-velocity features cluster at the tangent points $\phi = \pi/2$ on the receding (redshifted) side and $\phi = 3\pi/2$ on the approaching (blueshifted) side, fixing the sign of $V_{\rm los}$ in~\cref{eq:vz}.
The transformation from the disc frame to the sky frame is a rotation by $i$ about the disc major axis followed by a rotation by $\Omega$ about the LOS.
The black hole sky position $(x_0, y_0)$ relative to the VLBI phase centre is not directly observed and must be inferred from the maser spot positions.
Applying the rotation to a spot at $(r, \phi)$ and adding $(x_0, y_0)$ gives the predicted sky-plane coordinates:
\begin{align}
    x & = x_0 + r\bigl[\sin\phi\,\sin\Omega - \cos\phi\,\cos\Omega\,\cos i\bigr], \label{eq:pos_x} \\
    y & = y_0 + r\bigl[\sin\phi\,\cos\Omega + \cos\phi\,\sin\Omega\,\cos i\bigr]. \label{eq:pos_y}
\end{align}

The Keplerian circular velocity at orbital radius $r$ is
\begin{equation}\label{eq:vkep}
    V_\mathrm{kep}(r) = \Cv \sqrt{\frac{\MBH / \Msun}{(r / \mathrm{mas})\,(\DA / \Mpc)}},
\end{equation}
where $\Cv \approx 0.94~\kmsec$.\footnote{$\Cv = \sqrt{G\Msun / (1\,\mathrm{mas} \times 1\,\Mpc)} \times 10^{-3}$, the circular velocity $\sqrt{G\MBH/(r\DA)}$ expressed with $\MBH$ in $\Msun$, $r$ in mas, $\DA$ in Mpc, and $V_\mathrm{kep}$ in $\kmsec$.}
The model prediction for the observed velocity of a maser spot is the composition of three independent redshift contributions,
\begin{equation}\label{eq:vobs}
    V_i = c\Bigl[(1 + \vrec/c)(1 + z_D)(1 + z_g) - 1\Bigr].
\end{equation}
The dominant term is the recession redshift $\vrec/c$, encoding both the cosmological expansion and the galaxy's peculiar motion.
The recession velocity $\vrec = V_\mathrm{sys,obs} + \Delta V_\mathrm{sys}$ combines the observed CMB-frame systemic velocity (\cref{tab:data_availability}) with a sampled offset $\Delta V_\mathrm{sys}$ that accounts for the unknown systemic velocity of the disc relative to the catalogued host value.
The relativistic Doppler redshift from the orbital motion is
\begin{align}
    1 + z_D          & = \gamma\left(1 + \frac{V_{\rm los}}{c}\right), \label{eq:zD}                                           \\
    \gamma           & \equiv \frac{1}{\sqrt{1 - V_\mathrm{orb}^2/c^2}}, \label{eq:gamma}                                      \\
    V_{\rm los}      & = V_\mathrm{kep}\,\sin i\,\frac{\sin\phi + e\sin\omega}{\sqrt{1 + e\cos(\phi - \omega)}}, \label{eq:vz} \\
    V_\mathrm{orb}^2 & = V_\mathrm{kep}^2\,\frac{1 + e^2 + 2e\cos(\phi - \omega)}{1 + e\cos(\phi - \omega)}, \label{eq:vorb}
\end{align}
where $V_\mathrm{kep}(r)$ is the circular speed at the radius $r$, given by~\cref{eq:vkep}, $e$ is the orbital eccentricity, and $\omega(r) = \omega_0 + (\mathrm{d}\omega/\mathrm{d}r|_{r_\mathrm{ref}^{\omega}})(r - r_\mathrm{ref}^{\omega})$ is the warped argument of periapsis.
The circular limit $e \to 0$ recovers $V_{\rm los} = V_\mathrm{kep}\sin\phi\sin i$ and $V_\mathrm{orb} = V_\mathrm{kep}$.
The Lorentz factor $\gamma$ depends on the total orbital speed $V_\mathrm{orb}$, not on its LOS projection $V_{\rm los}$.
This produces a transverse Doppler redshift of order $\gamma - 1 \approx V_\mathrm{orb}^2 / 2c^2$ that persists even when the spot moves purely across the sky ($V_{\rm los} = 0$), as relativistic time dilation slows the emitting clock.
If the black hole potential is approximated as Schwarzschild, the gravitational redshift is
\begin{equation}\label{eq:zg}
    1 + z_g = \frac{1}{\sqrt{1 - R_\mathrm{s}/(r\DA)}}, \quad R_\mathrm{s} = \frac{2G\MBH}{c^2}.
\end{equation}
The centripetal acceleration projected along the LOS is
\begin{equation}\label{eq:accel}
    a_{\rm los} = \Ca\,\frac{\MBH}{r^2\,\DA^2}\,\cos\phi\,\sin i,
\end{equation}
where $\Ca = 1.872 \times 10^{-4}~\kmsecyr$.\footnote{$\Ca = G\Msun / (1\,\mathrm{mas} \times 1\,\Mpc)^2$, the centripetal acceleration $G\MBH/(r\DA)^2$ expressed with $\MBH$ in $\Msun$, $r$ in mas, $\DA$ in Mpc, and $a_\mathrm{los}$ in $\kmsecyr$.}
We assume the acceleration has the circular centripetal form, because the eccentricity corrections enter at higher order in $e$ and are negligible for the small eccentricities considered here.\footnote{We fix $e = 0$ in the models reported here.
The eccentric-disc variant tested in Section~\ref{sec:results_h0} returns eccentricities consistent with zero for every galaxy in the $H_0$ sample and leaves the distances and combined $H_0$ unchanged.
NGC~4258, which we exclude from that sample, is the exception: its eccentricity is offset from zero by $4.4\,\sigma$ (Section~\ref{sec:ngc4258}).}

\begin{figure*}
    \centering
    \includegraphics[width=\textwidth]{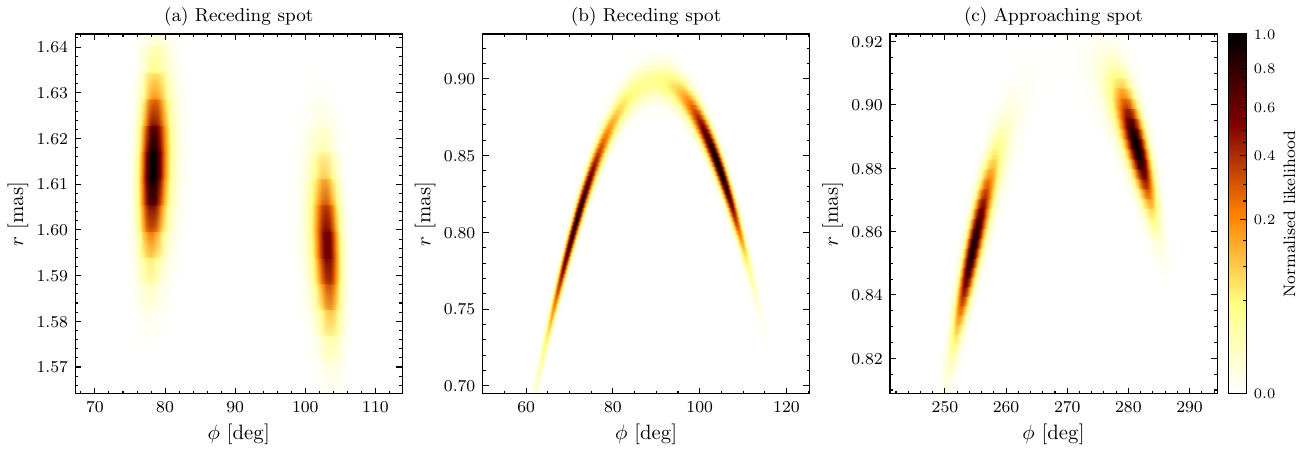}
    \caption{Illustration of the per-spot azimuthal bimodality in $\phi$ for NGC~5765b.
        Each panel shows the data likelihood $\mathcal{L}(\bm{d}_i \mid r,\, \phi,\, \boldsymbol{\theta})$ of a single maser spot as a function of its orbital angular radius $r$ and azimuthal angle $\phi$, with the disc parameters $\boldsymbol{\theta}$ fixed at the baseline linear-warp posterior median and the colour scale normalised to each spot's peak.
        The likelihood splits into the two solutions $\phi$ and $\pi - \phi$, symmetric about the tangent point $\phi = \pi/2$ for receding spots (panels a and b) and $\phi = 3\pi/2$ for approaching spots (panel c).
        In every panel the two solutions lie on the near and far sides of the disc along the LOS, so in the edge-on limit they project to the same sky position and share the LOS velocity ($V_{\rm los} \propto \sin\phi$;~\cref{eq:vz}) while having opposite LOS accelerations ($a_{\rm los} \propto \cos\phi$;~\cref{eq:accel}).
        The acceleration data only partially resolve this degeneracy, so the per-spot azimuthal angle remains bimodal and is marginalised over rather than fixed to a single solution.}
    \label{fig:rphi_bimodality}
\end{figure*}

In summary, the disc model maps the per-spot orbital coordinates $(r_i, \phi_i)$ and the disc parameters $\boldsymbol{\theta}$ to predicted sky positions through~\cref{eq:pos_x,eq:pos_y}, velocities through~\cref{eq:vobs}, and accelerations through~\cref{eq:accel}, all of which are compared with the observations in the likelihood below.
In a nearly edge-on disc ($i \approx 90^\circ$, $\cos i \approx 0$), the azimuthal angle enters the sky position and LOS velocity ($V_{\rm los} \propto \sin\phi\sin i$) only through $\sin\phi$, so spots at $\phi$ and $\pi - \phi$ share a position and velocity but have opposite accelerations ($a_{\rm los} \propto \cos\phi\sin i$).
The accelerations break this $\phi \leftrightarrow \pi - \phi$ degeneracy.
Spots at $\phi$ and $\phi + \pi$ instead have velocities and accelerations of opposite sign, a degeneracy broken by the velocity sign for the high-velocity features and by the acceleration sign for the systemic ones.
In practice the degeneracy is only partially broken, because accelerations are measured for only a subset of spots and have their own measurement uncertainty, so the per-spot likelihoods contain both azimuthal modes with varying amplitude ratios, as illustrated in~\cref{fig:rphi_bimodality}.

\subsection{Sampling}\label{sec:sampling}

For the $i$\textsuperscript{th} maser spot, we treat the position, velocity, and acceleration measurements as independent, so the per-spot likelihood factorises into three terms:
\begin{align}\label{eq:Lk}
     & \mathcal{L}(x_{\mathrm{obs}},\, y_{\mathrm{obs}},\, V_{\mathrm{obs}},\, a_{\mathrm{obs}} \mid r,\, \phi,\, \boldsymbol{\theta}) \nonumber \\
     & \quad = \mathcal{L}(x_{\mathrm{obs}},\, y_{\mathrm{obs}} \mid r,\, \phi,\, \boldsymbol{\theta})
    \mathcal{L}(V_{\mathrm{obs}} \mid r,\, \phi,\, \boldsymbol{\theta}) \nonumber                                                                \\
     & \qquad \times \mathcal{L}(a_{\mathrm{obs}} \mid r,\, \phi,\, \boldsymbol{\theta}),
\end{align}
where the spot index $i$ is suppressed for clarity.

The position likelihood is a product of independent Gaussians in $x$ and $y$, each with variance equal to the per-spot measurement uncertainty and an error floor ($\sigma_{x}$ and $\sigma_{y}$, respectively) added in quadrature,
\begin{align}\label{eq:L_pos}
     & \mathcal{L}(x_{\mathrm{obs}},\, y_{\mathrm{obs}} \mid r,\, \phi,\, \boldsymbol{\theta}) \nonumber               \\
     & \quad = \mathcal{N}\!\left(x_{\mathrm{obs}} \mid x,\; \sigma_{x,\mathrm{obs}}^2 + \sigma_{x}^2\right) \nonumber \\
     & \qquad \times \mathcal{N}\!\left(y_{\mathrm{obs}} \mid y,\; \sigma_{y,\mathrm{obs}}^2 + \sigma_{y}^2\right).
\end{align}
The velocity likelihood is
\begin{equation}\label{eq:L_vel}
    \mathcal{L}(V_{\mathrm{obs}} \mid r,\, \phi,\, \boldsymbol{\theta}) = \mathcal{N}\!\left(V_{\mathrm{obs}} \mid V,\; \sigma_{v,\mathrm{obs}}^2 + \sigma_v^2\right),
\end{equation}
where $\sigma_{v,\mathrm{obs}} = 0.25~\kmsec$ is the fixed per-spot velocity measurement floor and $\sigma_v$ is an error floor added in quadrature.
The error floor takes one of two values set by the spectral classification of the spot in the catalogue,
\begin{equation}\label{eq:sigma_v_class}
    \sigma_v = \begin{cases}
        \sigma_{v,\mathrm{sys}} & \text{systemic features},      \\
        \sigma_{v,\mathrm{hv}}  & \text{high-velocity features}.
    \end{cases}
\end{equation}
The two populations sample different disc regions and gain geometries: systemic features arise along the line of sight to the nucleus, where they amplify the background continuum, whereas high-velocity features arise near the tangent points, where long velocity-coherent paths along the orbit produce the strongest emission.
As in the MCP disc-modelling analyses, we therefore allow their non-thermal broadening to differ rather than impose a common value.
For NGC~5765b we split the systemic floor once more between the two velocity clumps, so that the clump~2 spots have their own position, velocity, and acceleration floors.

Not all spots have a measured acceleration, and coverage is partial in every galaxy (\cref{tab:data_availability}).
The spots that lack one are predominantly high-velocity features near the tangent points, where $a_{\rm los}$, which scales as $\cos\phi$ in~\cref{eq:accel}, is small and the multi-year spectral drift is undetectable.
We do not model this acceleration selection.
Spots with a measured acceleration contribute the Gaussian acceleration likelihood
\begin{equation}\label{eq:L_acc}
    \mathcal{L}(a_{\mathrm{obs},i} \mid r,\, \phi,\, \boldsymbol{\theta}) = \mathcal{N}\!\left(a_{\mathrm{obs},i} \mid a_{\rm los},\; \sigma_{a,i}^2 + \sigma_{a}^2\right),
\end{equation}
where $a_{\rm los} = a_{\rm los}(r, \phi, \boldsymbol{\theta})$ is the predicted acceleration given by~\cref{eq:accel}.

The per-galaxy posterior is proportional to the product over spots of the per-spot terms in~\cref{eq:Lk}, times the disc-parameter priors of~\cref{tab:priors}.
At stage~1 we sample the angular-diameter distance $\DA$ directly under a uniform prior, as in~\citetalias{Pesce2020}, with bounds obtained by mapping the comoving distance range implied by the observed host redshift through the fiducial cosmology.
The uniform-in-volume prior on the comoving distance is imposed later by reweighting the per-galaxy distance posteriors when inferring $H_0$ (Section~\ref{sec:joint_prob}).
The redshift likelihood relating $D$ to $H_0$ enters only at stage~2.

This posterior depends on the latent orbital coordinates $(r_i, \phi_i)$, which we treat differently in the two steps of the computation described below.
The posterior sampling draws them explicitly, jointly with the global disc parameters $\boldsymbol{\theta}$.
The differential-evolution search that seeds the chain instead marginalises them out, replacing each per-spot likelihood by its $(r, \phi)$-marginalised form
\begin{align}\label{eq:marginal_de}
     & \mathcal{L}(x_{\mathrm{obs}},\, y_{\mathrm{obs}},\, V_{\mathrm{obs}},\, a_{\mathrm{obs}} \mid \boldsymbol{\theta}) \nonumber                                                           \\
     & \quad = \iint \mathcal{L}(x_{\mathrm{obs}},\, y_{\mathrm{obs}},\, V_{\mathrm{obs}},\, a_{\mathrm{obs}} \mid r,\, \phi,\, \boldsymbol{\theta})\,\frac{\dd{\phi}}{2\pi}\,\pi(r)\,\dd{r},
\end{align}
integrating over the orbital radius with its flat, improper prior $\pi(r) = 1$ and over the azimuthal angle with its uniform prior on $[0, 2\pi]$.
The radius integral runs over $[r_{\min}, r_{\max}] = [R_{\min}, R_{\max}]/\DA$, with the physical range $R \in [0.01, 1.5]$~pc set conservatively wider than the observed radial extent of the spots.
The physical bounds are held fixed across the search, so the angular limits $[R_{\min}, R_{\max}]/\DA$ depend on the candidate $\DA$ rather than being frozen at a fiducial value.
These bounds are finite only because the differential-evolution search marginalises $r$ by numerical integration.
The posterior sampling instead draws $r$ explicitly and so uses the improper flat radius prior of~\cref{tab:priors}.
The product over spots of these marginal terms, times the disc-parameter prior $\pi(\boldsymbol{\theta})$, is a marginal posterior in the global disc parameters alone.
The double integral is evaluated by log-space trapezoidal quadrature on a dense $(r, \phi)$ grid.
The azimuthal grid is fixed at ${\sim}10^4$ nodes per spot, with a dense core resolving the $\phi = \pm\pi/2$ peaks of the high-velocity features to ${\sim}0.01^\circ$.
At fixed global parameters and radius, the azimuthal angle remains bimodal (\cref{fig:rphi_bimodality}), so the differential-evolution search marginalises over both modes on this fixed, dense grid.
The posterior sampling instead draws $\phi$ explicitly, moving between the modes with the dedicated reflection move of Section~\ref{sec:init}.
The likelihood marginalised over $\phi$ is by contrast unimodal in $r$, so the radial grid can instead be adaptive: for each spot a profile estimate $\hat{r}_i(\boldsymbol{\theta})$ locates this peak (Appendix~\ref{app:data_estimates}), around which a fine local grid is joined with a coarser global grid, totalling ${\sim}400$ nodes per spot.

We adopt a flat prior on the angular orbital radius $r_i$, the quantity that enters the likelihood, as in the MCP disc-modelling analyses.
A prior flat in the physical radius $R_i = r_i\,\DA$ would scale as $\pi(r_i) \propto \DA$, making the radii informative about the distance, so that an explicit model of the spots' radial selection function would likely be needed.
It would also couple every $R_i$ to the sampled $\DA$, opening a strongly correlated $r_i$--$\DA$ ridge in the posterior that the sampler would have to traverse.

\subsubsection{Initialisation and sampling}\label{sec:init}

The posterior has a complex geometry, with strong parameter correlations and the bimodal per-spot azimuthal angles described above.
Poor initialisation can trap samplers in low-probability regions, so we first locate the global posterior mode.

First, we obtain data-driven estimates of the angular-diameter distance $\DA$ and black hole mass for each galaxy (Appendix~\ref{app:data_estimates}), which provide the physically motivated starting points of the initial differential-evolution population.
Second, we search for the maximum a~posteriori (MAP) point of the global parameters using differential evolution~\citep{Storn1997}, evaluating the marginalised objective of~\cref{eq:marginal_de}.
Differential evolution is a population-based, gradient-free global optimiser: at each generation it mutates every candidate by adding a scaled difference of two other randomly chosen members to a third, recombines this mutant vector with the candidate by coordinate-wise crossover, and retains whichever of the two yields the higher marginalised posterior.
The population therefore contracts onto the dominant mode without requiring derivatives of the quadrature-based objective.
The initial working population of \num{2000} is split equally between data-driven seeds spread along the $D$--$\eta$ degeneracy ridge---random distances at fixed $\eta$, with the disc geometry jittered about its data-driven estimate---and the distinct points with the highest posterior from a $2^{14} = \num{16384}$-point scrambled Sobol sequence~\citep{Sobol1967} of the prior volume.
The population is evolved for up to \num{5000} generations with early stopping after \num{500} generations without improvement.
The No-U-Turn Sampler~\citep[NUTS;][]{Hoffman2014} is local, so without this seed it nearly always stalls in low-probability regions and then never reaches the global mode.
An alternative is to forgo the optimise-then-sample split and explore the prior volume directly with nested sampling.
This alternative recovers posteriors consistent with the chains seeded by differential evolution, confirming that differential evolution finds the dominant mode, but nested sampling is substantially more expensive than the optimise-then-sample split.
The mode at the global optimum contains so much more posterior mass than any other that the posterior cannot be regarded as a set of modes of comparable weight, so the split leaves no competitive alternative unexplored.

Third, we sample the per-galaxy posterior---the product of the per-spot likelihoods of~\cref{eq:Lk} and the disc-parameter priors---with a Metropolis-within-Gibbs scheme implemented in \texttt{BlackJAX}~\citep{blackjax2024}, initialised at the differential-evolution MAP point.
When we add the optional quadratic-warp block, we repeat the differential-evolution search over the extended parameter set, so each variant is initialised at its own MAP point.
The global disc parameters $\boldsymbol{\theta}$ are updated with NUTS in an unconstrained reparametrisation, using a dense mass matrix and a step size adapted during warm-up, with exact gradients from \texttt{JAX} automatic differentiation~\citep{Bradbury2021}.
Whereas the differential-evolution search marginalises over the reflected azimuthal modes, the posterior sampling draws the per-spot latents directly for speed, so the sampler must move between the modes.
Each spot's latent is the pair $(z_{r,i}, \phi_i)$ of the non-centred log-radius residual $z_{r,i} = \log(r_i/\hat{r}_i)$ and the azimuthal angle $\phi_i$, where $\hat{r}_i(\boldsymbol{\theta})$ is the conditional radius estimate of Appendix~\ref{app:data_estimates}.
A vectorised adaptive random-walk Metropolis sweep updates all spots at once, proposing a correlated Gaussian step in this plane for each spot.
A dedicated reflection move relocates each high-velocity spot between its two $\phi$ modes, jumping in both $z_{r,i}$ and $\phi_i$ rather than flipping $\phi$ alone.
Each global NUTS step is interleaved with $n_\mathrm{inner} = 30$ such per-spot sweeps, the inner iterations of the Metropolis-within-Gibbs scheme.
The per-spot proposal scales are adapted during warm-up towards an acceptance rate of $0.35$, and the reflection move is attempted for each high-velocity spot with probability $0.25$ per sweep.
Each chain uses \num{10000} warm-up iterations and \num{20000} sampling iterations.
We run ten chains and assess convergence with the split-$\hat{R}$ statistic~\citep{Gelman1992, Vehtari2021} and the effective sample size, requiring $\hat{R} < 1.01$ for the global disc parameters and every per-spot latent.
The differential-evolution search evaluates its \num{2000}-member population in parallel, and we run it on a GPU, whereas the posterior sampling is light enough to run on a laptop CPU.

\subsection{Outlier rejection}\label{sec:clipping}

The~\dsours\ tables of Section~\ref{sec:datasets} are constructed by applying our own outlier rejection to the~\dsall\ tables, an iterative fit-and-clip procedure that uses the disc model and optimiser described above.
In each round we find the maximum a~posteriori point of the global disc parameters by differential evolution, fix the global parameters there, and score every spot
by the largest absolute residual across its sky-position and velocity channels, each normalised by its measurement uncertainty and the inferred error floor added in quadrature, averaged over the spot's conditional posterior in the orbital coordinates $(r,\,\phi)$.
We remove every spot whose score exceeds $2.5\,\sigma$ and repeat until the retained set is unchanged.
In practice the criterion stabilises quickly: no spot is removed after the fourth round for any galaxy, and only NGC~6264 stabilises after a single round.
We verify that clipping at $2$ or $3\,\sigma$ instead leaves the inferred distances unchanged, so the result does not depend on the threshold.
Our criterion differs from the~\citetalias{Pesce2020} one, which clips at approximately $3\,\sigma$, mainly in averaging each spot's residual over its conditional posterior in $(r,\,\phi)$ rather than evaluating it at the best-fitting orbital coordinates.

\subsection{Inference of the Hubble constant}\label{sec:joint_prob}

We infer $H_0$ from the per-galaxy distance posteriors of Section~\ref{sec:sampling} together with the host recession velocities.
Each disc run returns a posterior on the comoving distance $D$, independent of $H_0$ and computed under a flat distance prior.
Like the~\citetalias{Pesce2020} analysis, this is a two-stage procedure.
Rather than re-evaluating the spot-level likelihood, we use a non-parametric, kernel-density representation of each galaxy's distance posterior.
The alternative---a single joint inference over all five galaxies, their distances coupled through the shared $H_0$---is ideal, but the coupling collapses the sampler's step size and makes it prohibitively expensive.\footnote{We verify on test cases that the two-stage result matches this full joint posterior over all galaxies and $H_0$, sampled directly with NUTS after marginalising the per-spot latents, albeit at far greater cost.}

We combine the $N_\mathrm{gal} = 5$ galaxies in a hierarchical model with shared parameters $\boldsymbol{\Lambda}$.
In the baseline analysis $\boldsymbol{\Lambda} = (H_0, \sigma_\mathrm{pec}, cz_\mathrm{lim}, cz_\mathrm{width})$ comprises the Hubble constant, the peculiar-velocity scatter, and the two redshift-selection parameters (below).
The variant analyses replace the selection term or add a coherent bulk-flow velocity $\Vext$.
Each galaxy's comoving distance $D_g$ is sampled jointly with $\boldsymbol{\Lambda}$, its Gaussian kernel-density distance posterior tabulated on a grid and linearly interpolated during sampling.

The observed CMB-frame systemic velocity gives $z_{\mathrm{obs},g} = V_{\mathrm{sys},g}/c$ (Section~\ref{sec:data}).
Given $D$ and $H_0$, the predicted cosmological redshift follows from the comoving distance--redshift relation,
\begin{equation}\label{eq:dC}
    D(z, H_0) = \frac{c}{H_0} \int_0^z \frac{\dd{z'}}{\sqrt{\Omega_{\rm m}(1+z')^3 + (1 - \Omega_{\rm m})}},
\end{equation}
assuming flat $\Lambda$CDM with $\Omega_{\rm m} = 0.315$.
The predicted recession velocity combines this cosmological redshift with a line-of-sight peculiar velocity $V_\mathrm{pec}$,
\begin{equation}\label{eq:cz_pred}
    1 + z_\mathrm{pred} = (1 + z_\mathrm{cosmo})(1 + V_\mathrm{pec}/c).
\end{equation}
We model the redshift residual as a Gaussian of width $\sigma_\mathrm{pec}$, so the likelihood is
\begin{equation}\label{eq:L_redshift}
    \mathcal{L}(z_{\mathrm{obs}} \mid D, \boldsymbol{\Lambda}) = \mathcal{N}\!\left(cz_{\mathrm{obs}} \;\middle|\; cz_\mathrm{pred},\; \sigma_\mathrm{pec}^2\right).
\end{equation}
In the simplest analysis $V_\mathrm{pec} = 0$, so $z_\mathrm{pred} = z_\mathrm{cosmo}$ and the scatter $\sigma_\mathrm{pec}$ alone describes the peculiar motion.
Beyond this case, we take $V_\mathrm{pec}$ from a reconstruction of the local peculiar-velocity field, evaluated at the galaxy's sky direction $\hat{\bm{n}}_g$ and sampled distance $D_g$.
We consider two reconstructions: the linear field of~\citetalias{Carrick2015}, computed from the \TWOMPP\ galaxy distribution~\citep{Lavaux2011} using linear theory, and the \Manticore\ field of~\citet{McAlpine2025}, a Bayesian field-level inference that reconstructs constrained initial conditions from the observed galaxy distribution and forward-models them non-linearly into the present-day density and velocity fields of the local volume.
The \Manticore\ inference gives a set of posterior samples of the local field rather than a single realisation, and we marginalise over them by averaging the likelihood over them, so the reconstruction uncertainty propagates into the inferred $H_0$.
The reconstructed line-of-sight velocity $V_\mathrm{recon}$ enters as
\begin{equation}\label{eq:vpec_recon}
    V_\mathrm{pec} = \beta\, V_\mathrm{recon}(\hat{\bm{n}}_g, D_g) + \hat{\bm{n}}_g \cdot \Vext,
\end{equation}
with $\beta$ rescaling the amplitude of the reconstructed velocity and $\Vext$ a residual coherent bulk flow that represents the velocity sourced by mass beyond the reconstructed volume.
Both reconstructions are tabulated on a comoving grid in $h^{-1}\,\Mpc$, so we convert the sampled distance in units of $\Mpc$ to the reconstruction's comoving units of $h^{-1}\,\Mpc$ using the sampled $H_0$.
We sample $\beta$ for the linear field under a Gaussian prior of $0.43\pm 0.02$~\citepalias{Carrick2015} because linear theory fixes the velocity only up to the degeneracy $\beta \equiv f/b$, the ratio of the growth rate to the galaxy bias.
The \Manticore\ field is computed for a fixed cosmology and predicts velocities at a definite physical amplitude, so we fix $\beta = 1$.
The five masers cannot constrain $\Vext$ on their own, so we adopt an informative Gaussian prior obtained by calibrating each reconstruction against the Cosmicflows-4 $W1$-band Tully--Fisher sample~\citep{Kourkchi2020}, following the procedure of~\citet{Stiskalek_2025_VFO}.
This prior is centred on $|\Vext| = 111 \pm 15~\kmsec$ towards Galactic $(\ell,\,b) = (310^\circ,\,-11^\circ)$ for the \Manticore\ field and $201 \pm 16~\kmsec$ towards $(306^\circ,\,-16^\circ)$ for the linear field.
Our baseline analysis adopts the \Manticore\ reconstruction~\citep{McAlpine2025, Stiskalek_2025_VFO}, retaining the no-peculiar-velocity and linear cases as variants.
Combining the distance posteriors with the redshift likelihood gives the population posterior,
\begin{align}\label{eq:stage2_posterior}
     & \mathcal{P}(\boldsymbol{\Lambda},\, \{D_g\} \mid \{z_{\mathrm{obs},g}\}) \propto \pi(\boldsymbol{\Lambda})\, Z(\boldsymbol{\Lambda})^{-N_\mathrm{gal}} \nonumber \\
     & \quad \times \prod_{g} D_g^2\, \mathcal{P}_g(D_g)\,\mathcal{L}(z_{\mathrm{obs},g} \mid D_g,\, \boldsymbol{\Lambda})\,S(V_{\mathrm{sys},g}),
\end{align}
where $\pi(\boldsymbol{\Lambda})$ is the prior on the shared parameters, $\mathcal{P}_g(D_g)$ the $g$\textsuperscript{th} galaxy's kernel-density distance posterior, $D_g^2$ reweights it to the uniform-in-volume prior, $S(V_{\mathrm{sys},g})$ the per-galaxy selection probability, and $Z(\boldsymbol{\Lambda})$ the per-galaxy detection probability normalising the sample (below).
We adopt $H_0 \sim \mathcal{U}(10, 200)~\kmsecMpc$, a Maxwell prior on $\sigma_\mathrm{pec}$ with mean $250~\kmsec$, and a uniform prior on each $D_g$.
We sample the population posterior with NUTS in \texttt{BlackJAX}.

The five galaxies are drawn from a flux-limited parent sample~\citep{Braatz1996,Braatz2004}: a water maser is detected only if its $22$~GHz line flux exceeds the survey's sensitivity limit, so its detection probability falls with distance.
We must therefore account for this distance-dependent selection.
The MCP sample results from a multi-stage pipeline: (1)~AGN identification from optical and infrared catalogues, (2)~$22$~GHz water-maser detection with the Green Bank Telescope (GBT)~\citep{Kuo2018,Kuo2020}, (3)~identification of the triple-peaked disc spectrum, (4)~concurrent VLBI imaging and multi-epoch GBT monitoring for spot positions and accelerations, and (5)~suitability assessment for the three-dimensional warped-disc model.
The net effect of this pipeline is an ill-characterised joint function of flux ($F \propto D^{-2}$), disc inclination, black hole mass, and the morphological quality of the disc~\citep{Zhu2011,vandenBosch2016}.
Because this pipeline cannot be reconstructed from the published information, we approximate the selection by a soft threshold.
We adopt the standard selection-corrected hierarchical posterior~\citep{Loredo2004, Kelly2007, Kelly2008, Hogg2010, ForemanMackey2014, Mandel2019, Vitale2022, Stiskalek2026a, Stiskalek2026b}: each detected galaxy enters with its selection probability $S$, and the sample is normalised by the per-galaxy detection probability $Z(\boldsymbol{\Lambda})$ raised to the power $-N_\mathrm{gal}$, as in~\cref{eq:stage2_posterior}.
As a simple gauge of the sensitivity to the selection strategy, we place this threshold on one of two variables, the redshift or the inferred distance, and compute $S$ and $Z(\boldsymbol{\Lambda})$ accordingly.
The baseline threshold acts on redshift,
\begin{equation}\label{eq:S_cz}
    S(cz) = \Phi\!\left(\frac{cz_\mathrm{lim} - cz}{cz_\mathrm{width}}\right),
\end{equation}
with $\Phi$ the standard normal cumulative distribution function and $cz_\mathrm{lim}, cz_\mathrm{width}$ sampled nuisance parameters.
Each galaxy enters at its observed $V_{\mathrm{sys},g}$, while the normalisation is the detection probability integrated over the comoving volume,
\begin{equation}\label{eq:Z_cz}
    Z(\boldsymbol{\Lambda}) = \int \Phi\!\biggl(\frac{cz_\mathrm{lim} - cz_\mathrm{pred}(\bm{r})}{\sqrt{\sigma_\mathrm{pec}^2 + cz_\mathrm{width}^2}}\biggr) \dd[3]{\bm{r}},
\end{equation}
where $\bm{r}$ is the comoving position and $cz_\mathrm{pred}(\bm{r})$ the recession velocity predicted there by~\cref{eq:cz_pred}, with any bulk flow or reconstruction folded into $V_\mathrm{pec}$.
The convolution with the peculiar-velocity scatter adds $\sigma_\mathrm{pec}$ in quadrature to $cz_\mathrm{width}$.
Because $cz_\mathrm{pred}$ depends on $H_0$, this normalisation does too, so the selection enters the $H_0$ inference rather than cancelling.
The alternative threshold acts on distance,
\begin{equation}\label{eq:S_D}
    S(D) = \Phi\!\left(\frac{D_\mathrm{lim} - D}{D_\mathrm{width}}\right),
\end{equation}
with $D_\mathrm{lim}, D_\mathrm{width}$ sampled nuisance parameters, so each galaxy enters at its sampled $D_g$ and the normalisation $Z(\boldsymbol{\Lambda}) = \int S(D)\, D^2\,\dd{D}$ is independent of $H_0$.

Distance is not an observable but a quantity we infer, so, unlike the baseline threshold on redshift, this selection does not act on measured data.
We therefore treat~\cref{eq:S_D} as an effective selection in the latent distance, included only as the alternative in this sensitivity test.
The true selection is more complex than either soft threshold, so we use the threshold as a coarse proxy and bracket its impact with a no-selection variant.
This term models only why a galaxy enters the sample, not the measurement of its individual maser spots, which the per-spot likelihood in~\cref{eq:Lk} already describes.
We do not model the inhomogeneous Malmquist bias~\citep{LandySzalay1992, Hudson1994, StraussWillick1995}, under which the distance prior would be proportional to the local intrinsic source density rather than scaling uniformly with volume as $D^2$.
The constrained reconstructions supply the present-day matter density along each line of sight, but mapping it to a source density requires the bias of maser hosts relative to matter, which five galaxies cannot constrain.
We drop this term and keep the volume prior, remaining agnostic about the host density field.

\subsection{Differences from the Megamaser Cosmology Project}\label{sec:mcp_diff}

Our forward model shares its physics with the MCP analyses, implementing the same warped Keplerian disc in \texttt{JAX}.
At matched disc parameters, orbital coordinates, and physical constants, it reproduces the per-spot likelihood of the \texttt{fit\_disk} code of~\citet{Reid2013} (M.~Reid, private communication) to machine precision, agreeing to a median difference of $5\times10^{-12}$ in log-likelihood.
This residual is the double-precision (\texttt{float64}) rounding floor: the relative machine epsilon is $\varepsilon\approx2\times10^{-16}$, which for log-likelihood magnitudes of order $10^{4}$ gives an absolute agreement at the ${\sim}10^{-12}$ level rather than at $\varepsilon$ itself.
The two codes therefore implement an identical likelihood, so any difference in the inferred distances must come from the input data, the priors, or the sampling rather than from the disc model.
Like~\citetalias{Pesce2020}, we infer $H_0$ in two stages, and the differences lie not in the disc likelihood but in how we model the sample and the per-spot latents, and in how we explore the per-galaxy disc posterior.
The outlier criterion is a further difference, as described in Section~\ref{sec:clipping}.

At the population stage we add an approximate model of the sample selection (Section~\ref{sec:joint_prob}).
We also impose a uniform-in-volume ($\propto D^2$) distance prior there, whereas at the population stage~\citetalias{Pesce2020} place a flat prior on each galaxy's latent recession velocity $v_i$ rather than on any distance coordinate, a choice that is not equivalent to a prior flat in $D$.
Changing the latent integration coordinate from $v_i$ to $\DA$ introduces the Jacobian $|\mathrm{d}v_i/\mathrm{d}\DA|$, which is ${\approx}H_0$ at these redshifts, so their measure includes an extra factor of $H_0$ per galaxy relative to a prior flat in distance.
Their stage~1 disc chains separately use a prior log-flat in $\DA$ rather than flat, so the density they use as the distance likelihood is itself weighted by $1/\DA$~\citep{Pesce2026}.
The remaining differences concern the per-galaxy disc posterior, in which the warped geometry and reflected azimuthal angles make locating the global mode the central difficulty when the sampler is initialised from a random prior draw.
We address this difficulty with two departures from the MCP analyses, in our choice of sampler and in our treatment of the per-spot latents:
\begin{enumerate}
    \item \emph{Sampler.}
    A differential-evolution global search locates the highest-probability posterior mode before any sampling.
    We then sample the posterior with a Metropolis-within-Gibbs scheme: NUTS for the global disc parameters and an adaptive random-walk sweep for the per-spot coordinates (Section~\ref{sec:init}).
    \item \emph{Per-spot latents.}
    At fixed disc parameters, each high-velocity spot admits two reflected azimuthal solutions of~\cref{eq:vz}, shown in~\cref{fig:rphi_bimodality}.
    A sampler that explores the full joint space with local proposals must cross this reflection to sample both solutions.
    We numerically marginalise the per-spot radii and azimuthal angles in the global search, which then runs over the $14$ disc parameters alone, and use, in the posterior sampling, a dedicated reflection move that relocates each high-velocity spot between its modes (Section~\ref{sec:init}).
\end{enumerate}

\section{Results}\label{sec:results}

We first present the disc posteriors inferred for each galaxy and compare the resulting distances with those of~\citetalias{Pesce2020} (Section~\ref{sec:results_disc}), then quantify the impact of the MCP reanalysis on those distances (Section~\ref{sec:origin_shift}), combine the five distances into a constraint on the Hubble constant and quantify its dependence on the sample selection and the peculiar-velocity reconstruction, together with the shift the MCP reanalysis produces (Section~\ref{sec:results_h0}), and finally report NGC~4258 separately (Section~\ref{sec:ngc4258}).
Throughout, every headline quantity is reported for both baseline inputs---the~\dsmcp and \dsours tables---while quantities computed from the \dsorig or \dsall tables appear only in~\cref{tab:distance_pesce_comparison,tab:h0_variants}.

\subsection{Per-galaxy disc posteriors}\label{sec:results_disc}

\begin{figure}
    \centering
    \includegraphics[width=\columnwidth]{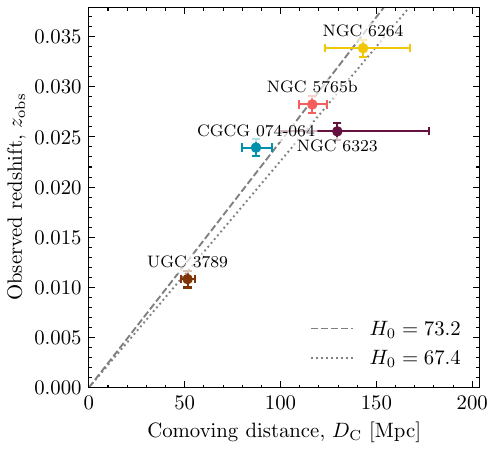}
    \caption{Observed CMB-frame redshift $z_\mathrm{obs}$ for each galaxy against its inferred comoving distance $D_\mathrm{C}$ on the~\dsours spot table (Section~\ref{sec:datasets}), with the horizontal error bars giving the central $68\%$ interval about the posterior median.
    The vertical error bars show a $250~\kmsec$ redshift uncertainty, representing the host peculiar velocity.
    The dashed and dotted lines are the flat-$\Lambda$CDM distance--redshift relations for the SH0ES distance-ladder ($H_0 = 73.2~\kmsecMpc$) and CMB ($H_0 = 67.4~\kmsecMpc$) values.
    }
    \label{fig:distance_redshift}
\end{figure}

On both baseline tables the five inferred distances agree with the~\citetalias{Pesce2020} angular-diameter distances: every median agrees to within $0.3\,\sigma$ on the~\dsmcp table, the largest absolute offset being NGC~6323 at $+13.8~\Mpc$ ($+0.29\,\sigma$), and within $0.4\,\sigma$ on the \dsours table.
Similarly, the two baselines agree with each other to within $0.3\,\sigma$ for every galaxy, the largest difference being UGC~3789 at $-1.9~\Mpc$ ($0.3\,\sigma$), so no distance depends appreciably on the outlier criterion or, for the NGC~5765b clump~2 spots, on whether they are clipped or instead retained and downweighted by a separate error floor.
The sample splits into three well-sampled discs (NGC~5765b, CGCG~074-064, and UGC~3789) and two others (NGC~6264 and NGC~6323), whose sparser maser spot coverage---in particular far fewer systemic features to anchor the disc geometry---leaves the distance posteriors far wider.
In~\cref{fig:distance_redshift} we plot the host recession redshifts against the inferred distances, which constrain $H_0$ once the sample is combined.

Our results here are stage~1 distances under a prior flat in $\DA$, whereas the~\citetalias{Pesce2020} stage~1 chains use a prior flat in $\log\DA$~\citep{Pesce2026}, so the posteriors are not expected to coincide exactly.
Reweighting a posterior of width $\sigma_D$ by $1/\DA$ lowers its median by ${\approx}\sigma_D^2/\DA$, which is ${\approx}11~\Mpc$ for NGC~6323 and below $3~\Mpc$ for the rest, so the prior accounts for the $+13.8~\Mpc$ offset of NGC~6323.
The remaining offsets span $-1.5$ to $+1.3~\Mpc$ and take both signs, whereas the reweighting can only raise our medians, so we cannot attribute them to the prior.
Neither the flat-in-$\DA$ nor the flat-in-$\log\DA$ prior is physically accurate: since galaxies are roughly uniformly distributed in comoving volume, the correct prior scales with comoving distance squared, which we apply in stage~2.

\begin{figure*}
    \centering
    \includegraphics[width=\textwidth]{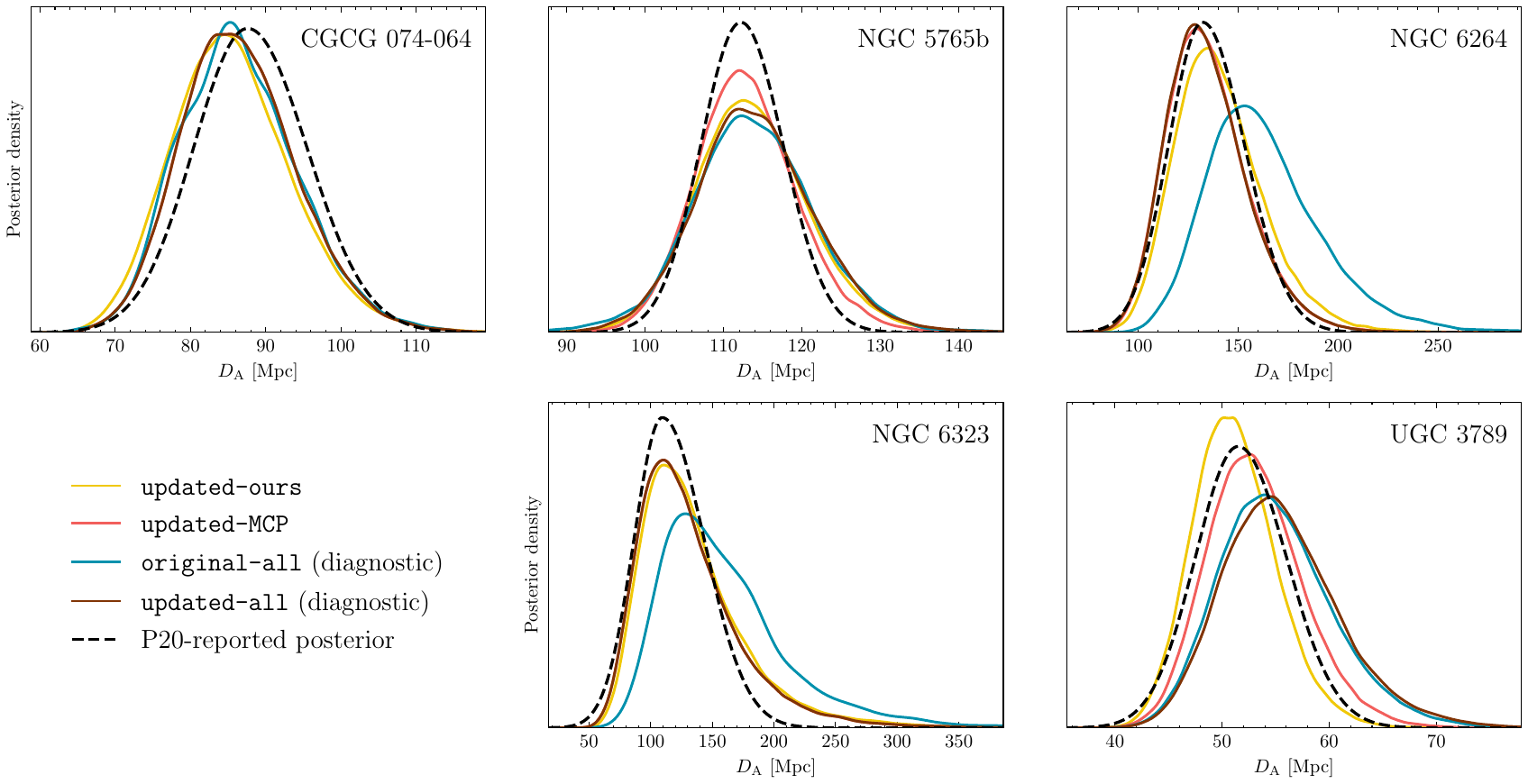}
    \caption{Angular-diameter distance posterior of each galaxy under the four spot tables of Section~\ref{sec:datasets}---\dsorig (blue), \dsmcp (red), \dsall (brown), and \dsours (yellow)---together with the~\citetalias{Pesce2020} posterior (black, dashed).
    The priors, sampler, and disc model are held fixed across the four inferences, except that the~\dsmcp model has no NGC~5765b clump~2 floors (Section~\ref{sec:disc_model}).
    The~\dsmcp and \dsours tables are based on the current MCP measurements and differ in the outlier treatment.}
    \label{fig:spot_table_comparison}
\end{figure*}

\Cref{tab:disc_params} reports the linear-warp posteriors on both baseline tables, with the quadratic-warp counterparts deferred to Appendix~\ref{app:qw}.
Both the nuisance parameters and the distances are stable between the two baselines, with one systematic exception: the error floors are larger on the \dsours table wherever it contains spots that the MCP cut removed.
For NGC~5765b, whose \dsours table has $206$ spots against $169$ in the~\dsmcp table, the sky-position floor $\sigma_x$ rises from $3.9^{+2.3}_{-1.9}$ to $6.0^{+2.2}_{-2.4}~\muas$ and the acceleration floor $\sigma_a$ from $0.051^{+0.021}_{-0.018}$ to $0.086^{+0.022}_{-0.019}~\kmsecyr$, even though the \dsours model additionally assigns the clump~2 spots their own floors.
The clump~2 floors of this galaxy are larger still, at $\sigma_x^{(2)} = 12.2^{+1.8}_{-2.0}~\muas$ and $\sigma_a^{(2)} = 0.15^{+0.07}_{-0.06}~\kmsecyr$, suggesting that the second clump indeed originates from a distinct, plausibly non-physical distribution.
An example of the full global-parameter posterior, for NGC~5765b, is shown in Appendix~\ref{app:corner}.
We infer these per-galaxy distances under the stage-1 flat prior on $\DA$, so they are not volume-corrected distance estimates and should not be used at face value.
Any downstream analysis that reuses them, such as a population-level $H_0$ inference, \emph{must} reapply both the uniform-in-volume ($\propto D^2$) distance prior and the selection function that we impose in stage~2 (Section~\ref{sec:joint_prob}), since omitting them biases the result~\citep{Desmond2026}.

The per-galaxy results are given in~\cref{tab:disc_params,tab:distance_pesce_comparison}.
On the~\dsmcp spot tables, only two distances change with the warp order: the quadratic warp raises NGC~5765b by $6.2~\Mpc$ ($0.56\,\sigma$) and NGC~6323 by $30.6~\Mpc$ ($0.49\,\sigma$), and leaves the other three within $2.5~\Mpc$ of their linear-warp medians.
The \dsours table gives the same picture for four of the five, moving CGCG~074-064, UGC~3789, and NGC~6264 by less than $2.7~\Mpc$ and NGC~5765b by $+6.0~\Mpc$ ($0.56\,\sigma$).
NGC~6323 is the exception: its $+30.6~\Mpc$ shift on the~\dsmcp table becomes $-1.6~\Mpc$ on the \dsours one.
Neither shift on the~\dsmcp table is a significant detection of curvature, since no curvature coefficient there is offset from zero by more than $2.9\,\sigma$, the strongest being the inclination term of UGC~3789 at $-11.5^{+3.6}_{-4.3}~\degmassq$ (Appendix~\ref{app:qw}).
On the other hand, on the \dsours baseline the NGC~5765b inclination curvature is $-6.5^{+0.9}_{-0.9}~\degmassq$, so how curved a disc appears depends on the spot table, although the two baselines differ both in which spots are retained and in the clump~2 floors, so neither choice alone accounts for it.
We retain the linear warp as the fiducial model on both baselines, because no distance moves appreciably with the warp order on either.
We further verify that orbital eccentricity does not affect these distances.

\begin{table*}
    \centering
    \caption{Per-galaxy disc parameters inferred under the linear warp on each of the two baseline spot tables, given as the posterior median and central $68\%$ interval.
    Warp rates are evaluated at the pivot radius $r_\mathrm{ref}$ of~\cref{eq:warp_i}, set to approximately the median spot radius.
    \textit{Upper block}: the~\dsmcp maser spot table. \textit{Lower block}: the \dsours table, which has the same updated spot measurements and differs in which spots are retained (Section~\ref{sec:datasets}).
    The $\sigma^{(2)}$ rows are the error floors of the NGC~5765b systemic clump~2.}
    \label{tab:disc_params}
    \setlength{\tabcolsep}{12pt}
    \renewcommand{\arraystretch}{1.3}
    \begin{tabular}{lccccc}
        \toprule
        Parameter                                                               & CGCG~074-064                    & NGC~5765b                       & UGC~3789                          & NGC~6264                        & NGC~6323                        \\
        \midrule
        \multicolumn{6}{l}{\textit{\dsmcp table}} \\
        \midrule
        $\DA$ (\Mpc)                                                       & $86.1^{+8.0}_{-7.1}$      & $112.6^{+6.6}_{-6.0}$     & $52.8^{+4.7}_{-4.1}$        & $133^{+22}_{-18}$         & $123^{+45}_{-29}$         \\
        $\log(\MBH/\Msun)$                                                & $7.376^{+0.039}_{-0.037}$ & $7.620^{+0.025}_{-0.024}$ & $7.087^{+0.037}_{-0.035}$   & $7.444^{+0.067}_{-0.063}$ & $7.06^{+0.14}_{-0.12}$    \\
        $i_0$ (\degunit)                                                       & $90.58^{+0.60}_{-0.59}$   & $84.52^{+0.20}_{-0.19}$   & $89.45^{+0.25}_{-0.24}$     & $90.42^{+0.40}_{-0.42}$   & $91.82^{+0.35}_{-0.36}$   \\
        $\mathrm{d}i/\mathrm{d}r|_{r_\mathrm{ref}}$ (\degmas)      & $-0.1^{+16.6}_{-4.0}$     & $12.23^{+0.59}_{-0.64}$   & $8.3^{+1.1}_{-1.0}$         & $-1.9^{+4.0}_{-4.4}$      & $-5.4^{+4.0}_{-4.9}$      \\
        $\Omega_0$ (\degunit)                                                  & $101.41^{+0.40}_{-0.41}$  & $146.65^{+0.11}_{-0.10}$  & $221.401^{+0.094}_{-0.093}$ & $94.26^{+0.17}_{-0.17}$   & $189.53^{+0.13}_{-0.13}$  \\
        $\mathrm{d}\Omega/\mathrm{d}r|_{r_\mathrm{ref}}$ (\degmas) & $5.8^{+2.8}_{-2.5}$       & $-3.29^{+0.23}_{-0.25}$   & $-1.65^{+0.59}_{-0.60}$     & $16.6^{+2.2}_{-2.2}$      & $12.8^{+1.3}_{-1.3}$      \\
        $x_0$ (\muas)                                                   & $1.1^{+1.0}_{-1.1}$       & $-43.7^{+1.5}_{-1.5}$     & $-401.48^{+0.94}_{-0.95}$   & $5.0^{+1.2}_{-1.2}$       & $16.03^{+0.99}_{-0.97}$   \\
        $y_0$ (\muas)                                                   & $6.2^{+2.8}_{-2.8}$       & $-99.6^{+2.1}_{-2.2}$     & $-461.5^{+1.0}_{-1.0}$      & $7.8^{+1.7}_{-1.7}$       & $7.1^{+2.5}_{-2.5}$       \\
        $\Delta V_\mathrm{sys}$ ($\kmsec$)                                & $-265.3^{+1.7}_{-1.9}$    & $-135.30^{+0.87}_{-0.83}$ & $15.00^{+0.82}_{-0.81}$     & $66.64^{+0.82}_{-0.79}$   & $191.4^{+1.6}_{-1.7}$     \\
        $\sigma_x$ (\muas)                                              & $1.03^{+0.63}_{-0.38}$    & $3.9^{+2.3}_{-1.9}$       & $4.5^{+1.0}_{-1.1}$         & $1.33^{+0.95}_{-0.59}$    & $3.5^{+1.0}_{-1.0}$       \\
        $\sigma_y$ (\muas)                                              & $15.0^{+2.5}_{-2.4}$      & $3.7^{+1.4}_{-1.5}$       & $6.2^{+1.3}_{-1.3}$         & $5.6^{+2.0}_{-2.0}$       & $4.1^{+2.5}_{-2.2}$       \\
        $\sigma_{v,\mathrm{sys}}$ ($\kmsec$)                              & $1.79^{+0.86}_{-0.82}$    & $1.70^{+0.74}_{-0.78}$    & $1.74^{+0.90}_{-0.85}$      & $1.64^{+0.89}_{-0.82}$    & $2.01^{+0.97}_{-0.92}$    \\
        $\sigma_{v,\mathrm{hv}}$ ($\kmsec$)                               & $2.08^{+0.91}_{-0.91}$    & $1.76^{+0.86}_{-0.87}$    & $1.79^{+0.95}_{-0.89}$      & $1.22^{+0.69}_{-0.61}$    & $1.78^{+0.89}_{-0.87}$    \\
        $\sigma_a$ ($\kmsecyr$)                                           & $0.23^{+0.11}_{-0.11}$    & $0.051^{+0.021}_{-0.018}$ & $0.264^{+0.062}_{-0.064}$   & $0.085^{+0.061}_{-0.043}$ & $0.185^{+0.097}_{-0.089}$ \\
        \midrule
        \multicolumn{6}{l}{\textit{\dsours table}} \\
        \midrule
        $\DA$ (\Mpc) & $85.2^{+8.3}_{-7.4}$ & $113.4^{+7.4}_{-6.7}$ & $51.0^{+4.0}_{-3.6}$ & $138^{+24}_{-19}$ & $126^{+46}_{-29}$ \\
        $\log(\MBH/\Msun)$ & $7.373^{+0.041}_{-0.040}$ & $7.625^{+0.027}_{-0.026}$ & $7.071^{+0.033}_{-0.032}$ & $7.463^{+0.069}_{-0.064}$ & $7.07^{+0.14}_{-0.12}$ \\
        $i_0$ (\degunit) & $90.25^{+0.37}_{-0.38}$ & $84.92^{+0.18}_{-0.19}$ & $89.34^{+0.22}_{-0.20}$ & $90.37^{+0.41}_{-0.43}$ & $91.73^{+0.29}_{-0.31}$ \\
        $\mathrm{d}i/\mathrm{d}r|_{r_\mathrm{ref}}$ (\degmas) & $-4.3^{+2.5}_{-3.5}$ & $10.20^{+0.77}_{-0.86}$ & $7.94^{+0.94}_{-0.87}$ & $-2.0^{+3.8}_{-4.1}$ & $-4.6^{+3.4}_{-4.2}$ \\
        $\Omega_0$ (\degunit) & $102.15^{+0.31}_{-0.32}$ & $146.80^{+0.11}_{-0.11}$ & $221.363^{+0.086}_{-0.084}$ & $94.24^{+0.17}_{-0.16}$ & $189.61^{+0.11}_{-0.11}$ \\
        $\mathrm{d}\Omega/\mathrm{d}r|_{r_\mathrm{ref}}$ (\degmas) & $3.2^{+2.1}_{-2.1}$ & $-3.65^{+0.26}_{-0.29}$ & $-1.65^{+0.56}_{-0.56}$ & $16.3^{+2.1}_{-2.1}$ & $12.9^{+1.1}_{-1.1}$ \\
        $x_0$ (\muas) & $1.20^{+0.92}_{-0.98}$ & $-44.9^{+1.8}_{-1.8}$ & $-401.36^{+0.86}_{-0.88}$ & $4.5^{+1.2}_{-1.2}$ & $15.53^{+0.90}_{-0.89}$ \\
        $y_0$ (\muas) & $4.7^{+2.1}_{-2.2}$ & $-99.4^{+2.5}_{-2.8}$ & $-461.33^{+0.96}_{-0.95}$ & $7.2^{+1.6}_{-1.7}$ & $6.6^{+2.5}_{-2.5}$ \\
        $\Delta V_\mathrm{sys}$ ($\kmsec$) & $-265.0^{+1.7}_{-1.8}$ & $-133.24^{+1.01}_{-0.99}$ & $14.57^{+0.79}_{-0.76}$ & $66.85^{+0.91}_{-0.89}$ & $191.1^{+1.6}_{-1.6}$ \\
        $\sigma_x$ (\muas) & $1.11^{+0.69}_{-0.43}$ & $6.0^{+2.2}_{-2.4}$ & $3.20^{+0.93}_{-1.00}$ & $1.57^{+1.23}_{-0.75}$ & $1.70^{+0.91}_{-0.77}$ \\
        $\sigma_y$ (\muas) & $3.4^{+2.6}_{-1.9}$ & $4.2^{+1.8}_{-1.7}$ & $4.8^{+1.2}_{-1.3}$ & $5.9^{+2.0}_{-1.9}$ & $3.4^{+2.2}_{-1.8}$ \\
        $\sigma_{v,\mathrm{sys}}$ ($\kmsec$) & $1.89^{+0.90}_{-0.87}$ & $1.80^{+0.83}_{-0.82}$ & $1.83^{+0.91}_{-0.88}$ & $1.67^{+0.89}_{-0.81}$ & $2.04^{+0.98}_{-0.95}$ \\
        $\sigma_{v,\mathrm{hv}}$ ($\kmsec$) & $2.08^{+0.93}_{-0.93}$ & $2.77^{+0.76}_{-0.82}$ & $1.70^{+0.94}_{-0.79}$ & $2.06^{+0.69}_{-0.72}$ & $1.80^{+0.94}_{-0.87}$ \\
        $\sigma_a$ ($\kmsecyr$) & $0.25^{+0.12}_{-0.12}$ & $0.086^{+0.022}_{-0.019}$ & $0.294^{+0.063}_{-0.061}$ & $0.126^{+0.056}_{-0.049}$ & $0.190^{+0.096}_{-0.091}$ \\
        $\sigma_x^{(2)}$ (\muas) & --- & $12.2^{+1.8}_{-2.0}$ & --- & --- & --- \\
        $\sigma_y^{(2)}$ (\muas) & --- & $7.6^{+3.4}_{-3.0}$ & --- & --- & --- \\
        $\sigma_{v,\mathrm{sys}}^{(2)}$ ($\kmsec$) & --- & $2.08^{+0.96}_{-0.94}$ & --- & --- & --- \\
        $\sigma_a^{(2)}$ ($\kmsecyr$) & --- & $0.148^{+0.072}_{-0.064}$ & --- & --- & --- \\
        \bottomrule
    \end{tabular}
\end{table*}

\subsection{Distance sensitivity to the spot measurements and clipping}\label{sec:origin_shift}

We propagate the four spot tables of Section~\ref{sec:datasets} through the identical disc model, priors, and sampler, to separate the effect of the updated measurements from that of the spot removal, and we compare the resulting distances in~\cref{tab:distance_pesce_comparison,fig:spot_table_comparison}.
Inferred from the \dsorig catalogues, the distances exceed the~\citetalias{Pesce2020}-reported values by up to $40~\Mpc$ for four of the five galaxies, with those four posteriors $16$ to $58$ per cent broader.
The \dsorig and \dsmcp tables differ in two ways at once (Section~\ref{sec:datasets}): a fraction of the position and acceleration measurements have been updated through an internal MCP reanalysis of the raw data, and a fraction of the spots have been removed as non-physical emission regions via an iterative fit-and-clip procedure.
The \dsall table separates the two, because it has the updated measurements while including every spot.
Relative to the \dsorig catalogues, only NGC~6264 and NGC~6323 move, both by ${\sim}26~\Mpc$, and both posteriors narrow.
For NGC~5765b and UGC~3789 the \dsall and \dsorig tables have the same values to numerical precision, and their medians reproduce to $0.2$ and $0.5~\Mpc$ (${\leq}0.1\,\sigma$), while CGCG~074-064 is unchanged by construction.
These three galaxies therefore bound the end-to-end reproducibility of the pipeline rather than measure any effect of the reanalysis.

Applying our own iterative clipping to the \dsall table then leaves the distance posterior unchanged to within $0.3\,\sigma$ for every galaxy except UGC~3789, whose median moves down by $4.1~\Mpc$ ($0.6\,\sigma$).
For NGC~6264 and UGC~3789 the difference between the \dsall and \dsmcp spot tables is the spot removal alone, so for these two galaxies the measurement update and the spot removal separate cleanly.
For NGC~6264 the spot removal has no effect on the inferred distance and the shift is driven by the updated acceleration uncertainties: inflating $33$ of them by a median factor of $3$, with no other change to the table, lowers the median distance from $159$ to $133~\Mpc$ and narrows the posterior.
For UGC~3789 the reverse holds: there is no measurement update at all, and the shift is driven by the spot removal, qualitatively consistent between the~\citetalias{Pesce2020} criterion and our own, which move the median down by $2.2$ and $4.1~\Mpc$, respectively.
For NGC~6323 the two effects do not separate: its \dsall table both updates the sky positions of $67$ of the original $68$ spots and adds $19$ first-epoch spots from~\citet{Kuo2011}, so its ${\sim}26~\Mpc$ shift cannot be attributed to the revised astrometry alone.
Separating them requires a fifth table, the published spots carrying the updated positions, which we do not construct here.

Finally, our two baselines use the same updated MCP position and acceleration measurements, and differ in which spots are retained and, for NGC~5765b, in the error-floor model (Section~\ref{sec:disc_model}), so the difference between them reflects those two choices together.
For NGC~5765b the difference in retained spots is larger than the clipping alone implies, because $20$ of the $43$ spots absent from the \dsmcp table are the systemic features whose published accelerations are placeholder values (Section~\ref{sec:data}): we retain these spots with their acceleration likelihood masked rather than clipping the spots.
The choice of the clipping criterion leaves the posteriors of CGCG~074-064, NGC~6264, and NGC~6323 unchanged.
For UGC~3789, where the criterion is the only difference between the baselines, it lowers the median by $1.9~\Mpc$ ($0.3\,\sigma$), and for NGC~5765b the criterion and the error-floor model together raise it by $0.7~\Mpc$ ($0.1\,\sigma$).
No galaxy therefore has a distance that depends appreciably on which outlier criterion is applied, or, for the NGC~5765b clump~2 spots, on whether they are clipped by the MCP or retained and downweighted by a separate error floor, under either the linear- or the quadratic-warp model.

\begin{table*}
    \centering
    \caption{Angular-diameter distance, in \Mpc, inferred from each of the four spot tables of Section~\ref{sec:datasets} under a prior flat in $\DA$, compared to the reported~\citetalias{Pesce2020} value.
    Entries give the posterior median and central $68\%$ interval.
    The~\citetalias{Pesce2020} column, from their tables~1 and~2, is for a linear warp.
    The four this-work columns are the spot tables of Section~\ref{sec:datasets}: the two baselines \dsmcp and \dsours, both built on the current MCP measurements, and the diagnostic \dsall and \dsorig catalogues which quantify the impact of the MCP reanalysis.
    Every galaxy agrees with~\citetalias{Pesce2020} to within $0.3\,\sigma$ on the~\dsmcp baseline spot table and $0.4\,\sigma$ on the \dsours one.}
    \label{tab:distance_pesce_comparison}
    \renewcommand{\arraystretch}{1.45}
    \setlength{\tabcolsep}{5pt}
    \begin{tabular*}{\textwidth}{@{\extracolsep{\fill}}lccccccc@{}}
        \toprule
        & \multicolumn{6}{c}{This work $\DA$} & \citetalias{Pesce2020} $\DA$ \\
        \cmidrule(lr){2-7}
        & \multicolumn{4}{c}{Linear warp} & \multicolumn{2}{c}{Quadratic warp} & \\
        \cmidrule(lr){2-5}\cmidrule(lr){6-7}
        Galaxy & \dsmcp & \dsours & \dsall & \dsorig & \dsmcp & \dsours & \\
        \midrule
        CGCG~074-064 & $86^{+8}_{-7}$ & $85^{+8}_{-7}$ & $86^{+8}_{-7}$ & $86^{+8}_{-7}$ & $86^{+8}_{-7}$ & $86^{+8}_{-7}$ & $88^{+8}_{-7}$ \\
        NGC~5765b & $113^{+7}_{-6}$ & $113^{+7}_{-7}$ & $114^{+8}_{-7}$ & $114^{+8}_{-7}$ & $119^{+9}_{-9}$ & $119^{+9}_{-8}$ & $112^{+5}_{-5}$ \\
        UGC~3789 & $53^{+5}_{-4}$ & $51^{+4}_{-4}$ & $55^{+6}_{-5}$ & $55^{+6}_{-5}$ & $51^{+5}_{-4}$ & $51^{+4}_{-4}$ & $52^{+5}_{-4}$ \\
        NGC~6264 & $133^{+22}_{-18}$ & $138^{+24}_{-19}$ & $133^{+22}_{-18}$ & $159^{+31}_{-23}$ & $135^{+22}_{-18}$ & $141^{+24}_{-19}$ & $132^{+21}_{-17}$ \\
        NGC~6323 & $123^{+45}_{-29}$ & $126^{+46}_{-29}$ & $123^{+45}_{-29}$ & $149^{+55}_{-35}$ & $154^{+58}_{-43}$ & $125^{+46}_{-29}$ & $109^{+34}_{-23}$ \\
        \bottomrule
    \end{tabular*}
\end{table*}

\subsection{Hubble constant inference}\label{sec:results_h0}

\begin{figure}
    \centering
    \includegraphics[width=\columnwidth]{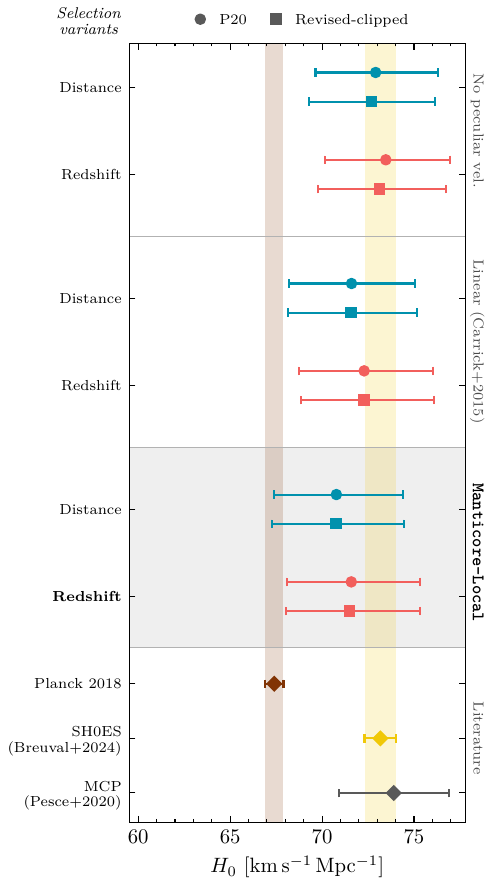}
    \caption{Inferred $H_0$ (median and central $68\%$ interval) over the reconstruction $\times$ selection grid, for the linear warp and the volume prior.
    Markers denote the baseline input spot table: filled circles the~\dsmcp table, filled squares the \dsours table (Section~\ref{sec:datasets}).
    Bands give the CMB~\citep{Planck2020cosmo} and SH0ES~\citep{Breuval2024} values for reference.}
    \label{fig:h0_measurement}
\end{figure}

Combining the five galaxy distances through the population model of~\cref{eq:stage2_posterior}, we find that the fiducial modelling configuration of Section~\ref{sec:joint_prob} (a redshift selection threshold, the \Manticore\ local-Universe reconstruction, the uniform-in-volume distance prior, and a linear warp) gives $H_0 = 71.6^{+3.8}_{-3.5}~\kmsecMpc$ on the~\dsmcp table and $71.5^{+3.9}_{-3.5}~\kmsecMpc$ on the \dsours one.
Both lie between the CMB value of~\citet{Planck2020cosmo} and the SH0ES distance-ladder value of~\citet{Breuval2024}, at $1.1\,\sigma$ above the former and $0.4\,\sigma$ below the latter, and neither distinguishes between them at the present precision.
Across all modelling variants, the two baseline sets of distances give $H_0$ values differing by at most $0.4~\kmsecMpc$ ($0.08\,\sigma$).
The two baselines therefore give the same $H_0$, though this joint constraint does not uniquely isolate the outlier clipping criterion, since the~\dsmcp model also has no NGC~5765b clump~2 floors: the two are alternative treatments of the same blended spots, the MCP clipping $12$ of the $28$ and we retaining and effectively downweighting them.
The two criteria share only $8$ of the ${\sim}31$ spots each removes, so $H_0$ is insensitive to which spots either of the two iterative clipping methods discards, but this does not bound a spot selection of a different kind.
The \dsorig catalogues give $69.3^{+3.7}_{-3.6}~\kmsecMpc$ in the same configuration, below the two baselines by ${\sim}2.3~\kmsecMpc$, an offset that persists across all modelling variants.
This offset combines the updated measurements, which contribute $1.8$ to $2.1~\kmsecMpc$ and originate mainly in NGC~6264 and NGC~6323, with the spot removal, which contributes the remainder.
The \dsorig results are therefore shown to quantify the impact of that reanalysis, and should not be interpreted as an alternative preferred analysis or as an additional systematic uncertainty.
\Cref{fig:h0_measurement,tab:h0_variants} compare the combined $H_0$ across the sample-selection threshold, the peculiar-velocity reconstruction, and the population-level distance prior, together with the four input tables.
Within either baseline spot table the eleven modelling variants span $2.4$ to $2.7~\kmsecMpc$.

The MCP sample is not volume-limited, so a distance-dependent selection must act.
We treat the distance- and redshift-selection runs as two effective, though not fully realistic, selection models, and the no-selection run (which unphysically omits the selection term entirely) as the control.
With the \Manticore\ reconstruction the two thresholds give $H_0 = 70.8^{+3.6}_{-3.4}$ (distance) and $71.6^{+3.8}_{-3.5}~\kmsecMpc$ (redshift) on the~\dsmcp tables, and $70.8^{+3.7}_{-3.5}$ and $71.5^{+3.9}_{-3.5}~\kmsecMpc$ on the \dsours tables.
The variable on which the threshold acts therefore moves $H_0$ by only $0.7$ to $0.8~\kmsecMpc$ on either baseline.
The selection parameters $cz_\mathrm{lim}$ and $cz_\mathrm{width}$ are only weakly constrained by the five galaxies (Appendix~\ref{app:population_corners}).
Unrealistically removing the selection entirely, at the volume prior and without a reconstruction, lowers $H_0$ by $2.2~\kmsecMpc$ on both baselines.

Relative to the no-peculiar-velocity case, the linear~\citetalias{Carrick2015} field lowers the median $H_0$ by $0.8$ to $1.4~\kmsecMpc$ and the non-linear \Manticore\ field of~\citet{McAlpine2025} by $1.6$ to $2.1~\kmsecMpc$, consistently across the four input tables.
The reconstruction runs also sample the residual bulk flow $\Vext$ of~\cref{eq:vpec_recon}, which the five galaxies do not constrain in any run, so its posterior recovers the informative prior calibrated against the Cosmicflows-4 sample.
\Cref{tab:vpec} gives the line-of-sight peculiar velocities each reconstruction assigns to the five hosts, averaged over their distance posteriors and separated into the reconstruction term and the $\Vext$ projection.
The reconstruction-only velocities of the two fields agree well, and the difference between the two fields arises mostly from the $\Vext$ projection.

Every variant in~\cref{tab:h0_variants} uses the linear warp, so we test the warp order separately on the \dsours table, with the \Manticore\ reconstruction and the volume prior.
Passing the quadratic-warp distances through the same population model gives $H_0 = 70.4 \pm 3.6~\kmsecMpc$ under the redshift selection and $69.4 \pm 3.5~\kmsecMpc$ under the distance selection, below their linear-warp counterparts by $1.1$ and $1.4~\kmsecMpc$ ($0.3$ and $0.4\,\sigma$).
This shift is driven primarily by NGC~5765b, whose distance the quadratic warp raises by $5.9~\Mpc$ ($5.2$ per cent).
The warp order therefore moves $H_0$ by less than half the statistical uncertainty of either run, so the inference is robust to this choice.

\begin{table*}
    \centering
    \caption{Combined $H_0$ in $\kmsecMpc$ (posterior median and central $68\%$ interval) for every modelling variant, on each of the four input spot tables (Section~\ref{sec:datasets}), all computed with the linear warp and the full five-galaxy sample.
    The \dsmcp and \dsours columns are the two baselines, the \dsall column separates the updated measurements from the spot clipping, except for NGC~6323, whose updated table also adds $19$ spots, and the \dsorig column uses the data as printed in the source papers, which the updated MCP measurements supersede.
    The \dsorig column quantifies the impact of that reanalysis and is not a preferred alternative result.
    The final column repeats each run using the~\citetalias{Pesce2020} stage~1 distance posteriors in place of ours, and agrees with the \dsmcp column.
    The fiducial configuration, in bold for both baselines, is the redshift selection with the \Manticore\ reconstruction and the volume prior.
    }
    \label{tab:h0_variants}
    \renewcommand{\arraystretch}{1.2}
    \setlength{\tabcolsep}{4pt}
    \small
    \begin{tabular*}{\textwidth}{@{\extracolsep{\fill}}lllccccc@{}}
        \toprule
        & & & \multicolumn{2}{c}{Baselines} & \multicolumn{2}{c}{Diagnostic} & Imported distances \\
        \cmidrule(lr){4-5}\cmidrule(lr){6-7}\cmidrule(lr){8-8}
        Selection & Reconstruction & Distance prior & \shortstack{\texttt{updated-}\\\texttt{MCP}} & \shortstack{\texttt{updated-}\\\texttt{ours}} & \shortstack{\texttt{updated-}\\\texttt{all}} & \shortstack{\texttt{original-}\\\texttt{all}} & \citetalias{Pesce2020} \\
        \midrule \midrule
        $-$ & $-$ & Volume & $71.2^{+3.3}_{-3.2}$ & $70.9^{+3.4}_{-3.3}$ & $70.8^{+3.0}_{-3.6}$ & $68.7^{+3.4}_{-3.3}$ & $71.4^{+3.1}_{-2.9}$ \\
        $-$ & $-$ & Distance & $72.5^{+3.3}_{-3.2}$ & $72.2^{+3.4}_{-3.3}$ & $71.9^{+3.6}_{-3.4}$ & $70.1^{+3.3}_{-3.3}$ & $72.5^{+3.1}_{-3.0}$ \\
        $-$ & $-$ & $\log \DA$ & $73.0^{+3.3}_{-3.2}$ & $72.8^{+3.5}_{-3.3}$ & $72.6^{+3.5}_{-3.4}$ & $70.8^{+3.5}_{-3.4}$ & $73.0^{+3.1}_{-3.0}$ \\
        $-$ & \citetalias{Carrick2015} & Distance & $71.1^{+3.6}_{-3.3}$ & $71.1^{+3.5}_{-3.4}$ & $70.6^{+3.7}_{-3.6}$ & $68.7^{+3.6}_{-3.3}$ & $71.2^{+3.3}_{-3.1}$ \\
        $-$ & \citetalias{Carrick2015} & $\log \DA$ & $71.7^{+3.6}_{-3.2}$ & $71.6^{+3.6}_{-3.2}$ & $71.2^{+3.7}_{-3.4}$ & $69.5^{+3.7}_{-3.4}$ & $71.7^{+3.3}_{-3.2}$ \\
        \midrule
        Distance & $-$ & Volume & $72.9^{+3.4}_{-3.3}$ & $72.7^{+3.4}_{-3.4}$ & $72.4^{+3.7}_{-3.4}$ & $70.6^{+3.6}_{-3.5}$ & $72.8^{+3.1}_{-3.0}$ \\
        Distance & \citetalias{Carrick2015} & Volume & $71.6^{+3.5}_{-3.4}$ & $71.6^{+3.6}_{-3.4}$ & $71.1^{+3.7}_{-3.6}$ & $69.2^{+3.6}_{-3.4}$ & $71.5^{+3.4}_{-3.2}$ \\
        Distance & \Manticore & Volume & $70.8^{+3.6}_{-3.4}$ & $70.8^{+3.7}_{-3.5}$ & $70.3^{+3.8}_{-3.6}$ & $68.5^{+3.7}_{-3.5}$ & $70.7^{+3.5}_{-3.3}$ \\
        \midrule
        Redshift & $-$ & Volume & $73.5^{+3.5}_{-3.3}$ & $73.1^{+3.6}_{-3.3}$ & $72.9^{+3.7}_{-3.4}$ & $71.1^{+3.6}_{-3.3}$ & $73.3^{+3.2}_{-3.1}$ \\
        Redshift & \citetalias{Carrick2015} & Volume & $72.3^{+3.8}_{-3.5}$ & $72.3^{+3.8}_{-3.4}$ & $71.9^{+3.9}_{-3.5}$ & $70.0^{+3.8}_{-3.4}$ & $72.2^{+3.5}_{-3.1}$ \\
        Redshift & \Manticore & Volume & $\mathbf{71.6^{+3.8}_{-3.5}}$ & $\mathbf{71.5^{+3.9}_{-3.5}}$ & $71.2^{+3.8}_{-3.6}$ & $69.3^{+3.7}_{-3.6}$ & $71.5^{+3.6}_{-3.4}$ \\
        \bottomrule
    \end{tabular*}
\end{table*}

\begin{table*}
    \centering
    \caption{Line-of-sight peculiar velocity of each maser host in the two reconstructions, in $\kmsec$ (median and central $68\%$ interval), averaged over the distance posterior of the baseline configuration on the~\dsmcp table and, for \Manticore, averaged over its realisations with equal weight.
    The velocity contribution is split into the reconstruction term $V_\mathrm{recon}$, rescaled by the velocity amplitude $\beta$, and the $\Vext$ projection, which together give $V_\mathrm{pec}$.
    We sample the velocity amplitude $\beta$ under a $\mathcal{N}(0.43, 0.02)$ prior for the linear field and fix $\beta = 1$ for \Manticore.
    $cz_\mathrm{obs}$ is the observed CMB-frame recession velocity of each host.}
    \label{tab:vpec}
    \renewcommand{\arraystretch}{1.2}
    \setlength{\tabcolsep}{4pt}
    \small
    \begin{tabular*}{\textwidth}{@{\extracolsep{\fill}}lccccccc@{}}
        \toprule
        & & \multicolumn{3}{c}{\citetalias{Carrick2015}} & \multicolumn{3}{c}{\Manticore} \\
        \cmidrule(lr){3-5}\cmidrule(lr){6-8}
        Galaxy & $cz_\mathrm{obs}$ & $V_\mathrm{recon}$ & $\hat{\bm{n}}_g \cdot \Vext$ & $V_\mathrm{pec}$ & $V_\mathrm{recon}$ & $\hat{\bm{n}}_g \cdot \Vext$ & $V_\mathrm{pec}$ \\
        \midrule \midrule
        CGCG~074-064 & $\num{7172}$ & $382^{+43}_{-201}$ & $9^{+13}_{-13}$ & $390^{+47}_{-202}$ & $362^{+70}_{-129}$ & $15^{+14}_{-14}$ & $376^{+73}_{-131}$ \\
        NGC~5765b & $\num{8526}$ & $184^{+165}_{-193}$ & $22^{+14}_{-14}$ & $205^{+165}_{-193}$ & $207^{+249}_{-184}$ & $22^{+14}_{-14}$ & $229^{+250}_{-185}$ \\
        UGC~3789 & $\num{3320}$ & $53^{+11}_{-18}$ & $-173^{+16}_{-16}$ & $-122^{+20}_{-23}$ & $53^{+39}_{-40}$ & $-97^{+15}_{-15}$ & $-45^{+42}_{-42}$ \\
        NGC~6264 & $\num{10193}$ & $232^{+96}_{-346}$ & $-67^{+16}_{-16}$ & $163^{+98}_{-347}$ & $243^{+123}_{-205}$ & $-28^{+15}_{-14}$ & $215^{+124}_{-207}$ \\
        NGC~6323 & $\num{7802}$ & $474^{+33}_{-103}$ & $-118^{+16}_{-16}$ & $354^{+38}_{-103}$ & $553^{+81}_{-158}$ & $-57^{+15}_{-14}$ & $495^{+83}_{-158}$ \\
        \bottomrule
    \end{tabular*}
\end{table*}

We now consider what the~\citetalias{Pesce2020}-reported distances give under our replication of their $H_0$ model.
Implementing their model with flat priors on $H_0$ and on each galaxy's latent recession velocity, the peculiar-velocity dispersion fixed at $250~\kmsec$, and their quoted statistical velocity errors, we infer $H_0 = 73.7 \pm 3.0~\kmsecMpc$.
Adding NGC~4258 through a Gaussian approximation to its published distance likelihood raises this by $0.26~\kmsecMpc$ to $73.9^{+3.1}_{-3.0}~\kmsecMpc$.
Both values reproduce their published results.
Passing the same distance posteriors through our own population model instead (the uniform-in-volume prior, sampled $\sigma_v$, the selection term, and the five-galaxy sample) but without any peculiar-velocity reconstruction gives $73.3^{+3.2}_{-3.1}~\kmsecMpc$.
With neither the selection term nor a reconstruction, our model gives $71.4^{+3.1}_{-2.9}~\kmsecMpc$ on the same distances.
Adding the \Manticore\ reconstruction to the run that includes the selection term lowers that value of $73.3~\kmsecMpc$ to $71.5^{+3.6}_{-3.4}~\kmsecMpc$, in agreement with our results when we infer the distances ourselves from the~\dsmcp tables.
For comparison,~\citetalias{Pesce2020} themselves report $71.8 \pm 2.9~\kmsecMpc$ when the host velocities are corrected with the~\citetalias{Carrick2015} flow field.
Our own model gives $72.2^{+3.5}_{-3.1}~\kmsecMpc$ on the same distances with that field, the redshift selection, and the uniform-in-volume distance prior.
However, the agreement in $H_0$ is not a like-for-like comparison, because the two population models differ in the ways listed above.
Most importantly, taking the recession velocity rather than the distance as the latent variable contributes an extra factor of $H_0$ per galaxy through the Jacobian (Section~\ref{sec:mcp_diff}).
This is the same $H_0$ dependence that a redshift-threshold selection introduces under a prior flat in distance, and that becomes $H_0^3$ per galaxy under a uniform-in-volume prior~\citep{Stiskalek2026a,Desmond2026}.
Their population model therefore already includes an approximate redshift selection accounting, which is why their value lies closer to the value from our selection run on the same distances ($73.3~\kmsecMpc$) than to that from the same run without the selection term ($71.4~\kmsecMpc$).

\subsection{NGC~4258}\label{sec:ngc4258}

NGC~4258 has little direct information about $H_0$, but its maser distance anchors the distance ladder, so we analyse it separately here.
We use the $358$-spot table described in Section~\ref{sec:data} and adopt the same modelling choices as in our stage~1 inference, including the prior flat in $\DA$, and model the disc alone with no selection function.\footnote{The distance prior is not important at this precision: reweighting the eccentric-disc posterior by $\DA^2$ would raise the median distance by $0.0015~\Mpc$ ($0.02\,\sigma$).}
Under the linear warp we infer $\DA = 7.416 \pm 0.056~\Mpc$, but the linear warp is decisively disfavoured for this galaxy.
The curvature coefficients are $\mathrm{d}^2i/\mathrm{d}r^2 = 0.22 \pm 0.02$ and $\mathrm{d}^2\Omega/\mathrm{d}r^2 = -0.14 \pm 0.02~\degmassq$.
The quadratic warp raises the distance to $\DA = 7.512 \pm 0.058~\Mpc$.
Allowing an eccentric disc, we infer $e = 0.0041 \pm 0.0009$ and the distance rises by $0.046~\Mpc$ ($0.49\,\sigma$) to $\DA = 7.558 \pm 0.073~\Mpc$.
Both distances are consistent with the $7.576 \pm 0.082\,\mathrm{(stat.)} \pm 0.076\,\mathrm{(sys.)}~\Mpc$ of~\citet{Reid2019}, at $0.51\,\sigma$ for the quadratic warp and $0.13\,\sigma$ for the eccentric disc, although our eccentricity lies $2.1\,\sigma$ below their $e = 0.007 \pm 0.001$.
Our uncertainties are statistical only.

\Cref{tab:ngc4258_variants} summarises the posteriors, and Appendix~\ref{app:ngc4258} shows the eccentric-model posterior.

\section{Discussion}\label{sec:discussion}

In this work, we revisit the megamaser $H_0$ inference, ranging from the maser disc modelling through the population-level combination to the peculiar-velocity treatment.
For the five MCP galaxies the fiducial configuration gives $H_0 = 71.5^{+3.9}_{-3.5}$ and $71.6^{+3.8}_{-3.5}~\kmsecMpc$ across the two baseline spot tables, a geometric, ladder-independent measurement in the local volume at $5$ per cent precision.

\subsection{Comparison to other megamaser $H_0$ measurements}

\citet{Boruah2021} use the same~\citetalias{Pesce2020} distances but account for peculiar velocities with the~\citetalias{Carrick2015} field, marginalising over the line-of-sight peculiar velocity, obtaining $H_0 = 70.1 \pm 2.9~\kmsecMpc$ under their fiducial uniform distance prior and $69.0^{+2.9}_{-2.8}~\kmsecMpc$ under a uniform-in-volume prior.
When reproducing their inference, we find that the offset from~\citetalias{Pesce2020} is not driven by the marginalisation itself: marginalising the line-of-sight velocity rather than evaluating it at a point moves the median by ${\lesssim}0.1~\kmsecMpc$.
The shift instead splits into ${\sim}2~\kmsecMpc$ from applying a~\citetalias{Carrick2015} correction at all, ${\sim}0.6~\kmsecMpc$ from replacing the~\citetalias{Pesce2020} flat prior on the latent recession velocity with one flat in distance, and ${\sim}1~\kmsecMpc$ from where the field is queried.
\citetalias{Pesce2020} evaluate it at the observed redshift, placing each host at its redshift-space position, whereas~\citet{Boruah2021} and we evaluate it in real space, at the sampled distance.
Adopting the redshift-space convention recovers their point-estimate value to $0.06~\kmsecMpc$, whereas every real-space evaluation---at the maser distance, iterated to self-consistency, or marginalised---lies ${\sim}1.2~\kmsecMpc$ below it.
Neither inference of~\citet{Boruah2021} includes a selection term: their fiducial prior uniform in distance is unphysical, while the uniform-in-volume variant particularly requires one, since without it a five-object realisation of that prior would place the hosts at substantially larger distances than observed.

\citet{Barua2025} inferred $H_0$ on inputs identical to those of~\citetalias{Pesce2020}, the only difference being that they profile the per-galaxy latent velocities rather than marginalise them, and recover $H_0 = 73.5^{+3.0}_{-2.9}~\kmsecMpc$ in agreement with~\citetalias{Pesce2020}.
This is not an independent measurement but simply the profile--marginal agreement expected for a near-Gaussian likelihood, and profiling is in any case an unprincipled choice as latent distances ought to be marginalised.

\citet{Watkins2026} reanalyse the~\citetalias{Pesce2020} distances with the \Manticore\ reconstruction and report $H_0 = 68.8 \pm 2.6~\kmsecMpc$, which is $2.7~\kmsecMpc$ below our fiducial value on the same distances and the same reconstruction.
Their model has $H_0$ as the only free parameter, with $\sigma_v$ fixed by hand and no latent distances or selection term.
Reimplementing their model, we recover every entry of their table~2 to $0.07~\kmsecMpc$.
When the velocity field is removed, their model gives a median of $73.4~\kmsecMpc$, so their velocity treatment lowers $H_0$ by $4.6~\kmsecMpc$.

The difference is the radius at which the reconstruction is queried.
Interpolating the \Manticore\ realisations along the six lines of sight, we recover the peculiar velocities of their table~1 to the quoted precision when the field is read at $0.681\,\DA$ in $h^{-1}\,\Mpc$, that is at the~\citetalias{Pesce2020} angular-diameter distance converted to $h^{-1}\,\Mpc$ with the $h$ of the reconstruction box rather than with the sampled $H_0$.
The conversion is implicit in how they set the cell size on which the velocity field is defined: they set it to $\num{1000}/256~\Mpc$, whereas the $256^3$ mesh spans $681.1~h^{-1}\,\Mpc$, a side of $\num{1000}~\Mpc$ only if $h = 0.681$.
The reconstruction is inferred from a redshift survey, so its radial coordinate, in $h^{-1}\,\Mpc$, is set by the observed redshifts and is independent of $h$.
The lookup distance for a host at comoving distance $D$ is therefore $r = h\,D$, evaluated at the sampled $h$,%
\footnote{The box radial coordinate $r$, in $h^{-1}\,\Mpc$, is a redshift coordinate if peculiar velocities are neglected: a structure observed at redshift $z$ is placed at $r = h\,D(z, 100\,h~\kmsecMpc)$, in which $h$ cancels by~\cref{eq:dC}.
The fiducial $h$ also enters the reconstruction's prior power spectrum, which has a negligible effect on the predicted velocities.}
whereas converting with the fixed $h = 0.681$ places every host at the radius it would occupy only if $H_0 = 68.1~\kmsecMpc$, and using angular diameter rather than comoving distance moves it inwards by a further factor $(1+z)$.
They likely adopt $\DA$ because it is the distance the maser geometry delivers directly and the quantity tabulated by~\citetalias{Pesce2020}, yet the comoving distance is required here because the reconstruction is defined on a comoving grid, so the field must be queried in the coordinate in which it is inferred.
For NGC~5765b this raises the assigned velocity by ${\sim}400~\kmsec$ on the host that contributes $55$ per cent of the Fisher information on $H_0$ in their model.

Starting from their $68.8~\kmsecMpc$: restricting the sample to the five galaxies we analyse and adopting $\sigma_v = 250~\kmsec$ raises $H_0$ to $69.4~\kmsecMpc$.
Replacing the angular-diameter by the comoving distance adds $0.5~\kmsecMpc$, and performing the conversion to $h^{-1}\,\Mpc$ with the sampled $H_0$ rather than with $h = 0.681$ adds a further $1.5~\kmsecMpc$, giving $71.4~\kmsecMpc$.
Fixing the lookup radius also sharpens the constraint artificially: letting it track the sampled $H_0$ widens the interval from $^{+3.0}_{-2.8}$ to $^{+3.8}_{-3.7}~\kmsecMpc$.
Adding the external bulk flow $\Vext$, which their model omits, shifts $H_0$ by a further ${\sim}0.1~\kmsecMpc$.
Sampling $\sigma_v$ under our Maxwell prior of mean $250~\kmsec$ rather than fixing it there moves $H_0$ by less than $0.05~\kmsecMpc$.
The tension they report is therefore an artefact of where the reconstruction is read rather than a genuine re-evaluation of the megamaser $H_0$.

\subsection{Caveats and future work}

We note five caveats.
First, our selection model is a coarse soft threshold placed on a single variable, whereas the true selection is an ill-characterised joint function of flux, inclination, black hole mass, and disc quality.
For the present sample the choice of selection model moves $H_0$ by less than a single model's statistical uncertainty, but for the larger samples anticipated from future maser surveys this choice would become the dominant systematic~\citep{Desmond2026}.
We place the threshold on either side of Hubble's law---on the observed redshift or on the inferred distance---and the two give $H_0$ values that differ by $0.7$ to $0.8~\kmsecMpc$.
This indicates that the inference is insensitive to which variable the threshold acts on, but it cannot bound the effect of the true selection, which is likely a joint function of several quantities, some of which we do not model.

Second, we retain the linear warp as the fiducial model because no distance moves by more than $0.6\,\sigma$ with the warp order on either baseline.
A linear warp is unlikely to be the physical form of the accretion disc, but where the curvature coefficients are consistent with zero the data do not constrain the curvature, and where they are not, the distance posterior is nonetheless little changed, so the warp order does not affect the inferred distance in either case.

Third, we keep a uniform-in-volume distance prior rather than also modelling the inhomogeneous Malmquist bias, because mapping the reconstructed matter density to a maser-host source density requires a host bias that five galaxies cannot constrain (see our treatment in~\citealt{Stiskalek2026a,Stiskalek_2025_VFO}).

Fourth, the analysis is conditional on the current MCP reduction of the interferometric data, which we adopt, and on an outlier assessment.
Moving from the published catalogues to the updated ones shifts $H_0$ by $1.8$ to $2.1~\kmsecMpc$ between the \dsorig and \dsall tables.
This shift arises entirely from NGC~6264, whose acceleration uncertainties are inflated, and NGC~6323, whose sky positions are revised and whose table is augmented with $19$ earlier-epoch spots.
The dependence on the outlier clipping is smaller: the \dsall and \dsours tables share the updated measurements and the NGC~5765b clump~2 floors and differ only in the $32$ spots we remove, and $H_0$ moves by at most $0.5~\kmsecMpc$ across the eleven modelling variants.
We adopt separate error floors for the clump~2 spots in NGC~5765b, which is effectively an outlier treatment: without it, the inferred distance to NGC~5765b would be higher by about $1\,\sigma$ and would propagate to an $H_0$ constraint lower by ${\lesssim}0.5\,\sigma$.
The MCP instead clip some of the clump~2 spots, and we show that the two approaches---separate error floors or clipping---give consistent results.
Removing that dependence requires forward-modelling closer to the raw data, along the lines of the epoch-by-epoch component tracking that~\citet{vanderBoom2025} apply to NGC~4258, so that spot identification and outlier rejection enter the likelihood rather than precede it.
Short of that, the hard $\sigma$ clip could be replaced by a Bayesian outlier model in which each spot has an outlier probability that is marginalised over.
This would allow the uncertainty in the outlier population to be correctly propagated.

Fifth, we model which galaxies enter the sample but not which spots are detected within a disc.
Maser amplification depends on the path length through the disc, so spots near the midline are preferentially detected, and the reported spots are therefore not a uniform sample of the disc.
Our (and the MCP) likelihood conditions on the spots that are reported and includes no per-spot detection probability, and may thus lead to a bias in the inferred disc geometry and distance.
Quantifying this bias would require a per-spot detection model, which we leave to future work.

\section{Conclusion}\label{sec:conclusion}

We have performed a fundamental reanalysis of the Hubble constant inference from megamasers, focused on testing its sensitivity to the warped Keplerian disc model and its posterior exploration, the sample selection, distance priors and peculiar-velocity modelling, and on quantifying the impact of the MCP reanalysis of the spot tables.
We adopt the physically motivated uniform-in-volume distance prior coupled to a self-consistent treatment of the selection function.
Absent knowledge of the sample's true selection, we adopt two simple models---distance and redshift selection---to gauge the effect of the selection treatment.

Our main findings are as follows:
\begin{enumerate}
    \item The five megamaser discs of the Megamaser Cosmology Project give $H_0 = 71.5$ and $71.6~\kmsecMpc$ ($\pm 3.6~\kmsecMpc$ statistical) on our two baseline spot tables, in the fiducial configuration of a redshift selection threshold, the \Manticore\ reconstruction, and a linear warp.
    Both values are consistent with the CMB ($1.1\,\sigma$ above) and the SH0ES distance ladder ($0.4\,\sigma$ below).
    \item Our disc modelling reproduces the~\citetalias{Pesce2020} distances to within $0.3\,\sigma$ when given their inputs.
    \item The MCP reanalysis of the spot tables, which supersedes the published catalogues, raises $H_0$ by $2.2$ to $2.5~\kmsecMpc$ at fixed modelling configuration (Section~\ref{sec:datasets}).
    Of this, $1.8$ to $2.1~\kmsecMpc$ comes from the updated measurements, which affect only NGC~6264 and NGC~6323, and the remainder from the removed spots.
    The as-published \dsorig catalogues give $69.3^{+3.7}_{-3.6}~\kmsecMpc$, reflecting inferred distances that exceed those in~\citetalias{Pesce2020} by up to $40~\Mpc$.
    The two largest offsets, $27$ and $40~\Mpc$, are those of NGC~6264 and NGC~6323, while the smaller ones arise from the spot removal and the stage-1 distance prior rather than from the reanalysis.
    We quote this to document the impact of that reanalysis, not to offer an alternative result.
    Our baseline is the current~\dsmcp and \dsours tables throughout.
    \item Within either of our baseline spot tables, the eleven modelling variants span only $2.4$ to $2.7~\kmsecMpc$, less than the statistical uncertainty of any one run.
    The distance prior has a larger effect than the selection modelling or velocity reconstruction; using a physical uniform-in-volume prior rather than the $\log(D_A)$ prior of~\citet{Pesce2020} reduces $H_0$ by 1.6--2.1 $\kmsecMpc$ across the dataset variants in the case of no selection model or peculiar velocity field (Table~\ref{tab:h0_variants}).
    \item Modelling the local peculiar-velocity field lowers $H_0$ relative to the no-reconstruction case, by $0.8$ to $1.4~\kmsecMpc$ for the linear~\citetalias{Carrick2015} field and by $1.6$ to $2.1~\kmsecMpc$ for the non-linear \Manticore\ field, consistently across the four input spot tables.
    \item A quadratic warp on the \dsours table lowers $H_0$ by $1.1$ to $1.4~\kmsecMpc$, less than half the statistical uncertainty. This shift is driven by NGC~5765b whose distance rises by $5.2$ per cent.
    \item For NGC~4258 (which we exclude from the $H_0$ inference), the eccentric quadratic-warp model gives $\DA = 7.558 \pm 0.073~\Mpc$, in good agreement ($0.13\,\sigma$) with the $7.576 \pm 0.082\,\mathrm{(stat.)} \pm 0.076\,\mathrm{(sys.)}~\Mpc$ of~\citet{Reid2019}.
\end{enumerate}

At $5$ per cent precision, the megamaser data does not meaningfully discriminate between the \textit{Planck} and SH0ES Hubble constant values. This will require
a larger sample of megamaser discs with complete spot and acceleration catalogues; the improved $22$~GHz sensitivity of the ngVLA is forecast to yield the additional systems needed to reach the ${\sim}1$ per cent level in $H_0$~\citep{Braatz2018}, at which point masers will become an independent, purely-geometric arbiter of the Hubble tension.

\begin{acknowledgments}
We thank Dominic Pesce and the MCP team for providing the spot tables used in the~\citetalias{Pesce2020} disc modelling, for clarifying how they were assembled, and for extensive discussion of the comparison presented here.
We thank Mark Reid for providing the \texttt{fit\_disk} code and for useful discussions.
We thank Pedro Ferreira, Guilhem Lavaux, and Adam Riess for helpful discussions of this work.
RS acknowledges financial support from a Hintze Fellowship at the Oxford Centre for Astrophysical Surveys, funded through generous support from the Hintze Family Charitable Foundation.
HD is supported by Royal Society University Research Fellowship grant 211046.
\end{acknowledgments}

\bibliographystyle{apsrev4-1}
\bibliography{ref}

\appendix

\section{Data-driven estimates of distance, mass, and orbital radius}\label{app:data_estimates}

We derive inexpensive, approximate estimates of the comoving distance, black hole mass, and per-spot orbital radii directly from the spot data, without sampling the full disc model.
They seed the differential-evolution search of Section~\ref{sec:init}: the distance and mass estimates set the $D$--$\eta$ ridge along which half the population is drawn, while the per-spot radius estimate $\hat{r}_i(\boldsymbol{\theta})$ centres both the conditional radius grid of the marginalised likelihood in~\cref{eq:marginal_de} and the non-centred radius reparametrisation $z_{r,i} = \log(r_i/\hat{r}_i)$ that the posterior sampling draws in place of $r_i$.

\paragraph{Comoving distance.}
From the CMB-frame recession velocity $V_\mathrm{sys}$, we estimate the comoving distance by the Hubble law with a second-order cosmographic correction,
\begin{equation}\label{eq:Dc_est}
    D_\mathrm{est} = \frac{cz_\mathrm{est}}{H_0}\left[1 - \tfrac{1}{2}(1 + q_0)\,z_\mathrm{est}\right],
\end{equation}
where $z_\mathrm{est} = V_\mathrm{sys}/c$, $q_0 = -0.5275$ for $\Omega_{\rm m} = 0.315$, and $H_0 = 73~\kmsecMpc$.
The angular-diameter distance estimate is $\DA^\mathrm{est} = D_\mathrm{est}/(1 + z_\mathrm{est})$.

\paragraph{Black hole mass.}
For each high-velocity spot $i$ at angular separation $r_i = \sqrt{x_i^2 + y_i^2}$ and velocity offset $\Delta V_i = |V_i - V_\mathrm{sys}|$, the Keplerian relation in~\cref{eq:vkep} gives
\begin{equation}\label{eq:MBH_est}
    M_i = \frac{\Delta V_i^2\, r_i\, \DA^\mathrm{est}}{\Cv^2},
\end{equation}
and we adopt $\log\MBH^\mathrm{est} = \mathrm{median}_i(\log M_i)$.
Neglecting the $\sin i$ projection biases this low by a factor of $\sin^2 i$, ${\lesssim}0.01$~dex for the near-edge-on maser discs, which is small compared with the prior width.

\paragraph{Per-spot orbital radius.}
Within the inference, each spot's radius is estimated from the Keplerian relations at its dominant azimuthal angle $\phi$, conditional on the current disc parameters $\boldsymbol{\theta}$.
For a high-velocity spot, near $\phi = \pm\pi/2$, we invert the LOS velocity in~\cref{eq:vz} to give $\hat{r} = \MBH\,(\Cv\sin i)^2/(\DA\,\Delta V^2)$.
For a systemic spot, at $\phi \approx 0$, we invert the LOS acceleration in~\cref{eq:accel} to give $\hat{r} = \sqrt{\Ca\,\MBH\sin i/(\DA^2\,|a_\mathrm{obs}|)}$.
The analytic seed is refined by a one-dimensional optimisation over $\log r$ of the $\phi$ integral in~\cref{eq:marginal_de}, using Brent's method~\citep{Brent1973}.

\section{Quadratic-warp disc posteriors}\label{app:qw}

\Cref{tab:disc_params_qw} reports the per-galaxy disc posteriors under the quadratic-warp model, the counterpart of the linear-warp results in~\cref{tab:disc_params}.
On the \dsours table, the NGC~5765b inclination curvature is $-6.5^{+0.9}_{-0.9}~\degmassq$, offset from zero by $7\,\sigma$, a stronger detection of curvature than any coefficient on the~\dsmcp table (Section~\ref{sec:results_disc}).
The two tables differ in which spots are retained and in the clump~2 floors (Section~\ref{sec:disc_model}), so the preference for curvature in that disc cannot be attributed to either choice alone.
The quadratic warp raises the NGC~5765b distance by $6.0~\Mpc$ ($0.56\,\sigma$) on the \dsours table and by $6.2~\Mpc$ ($0.55\,\sigma$) on the~\dsmcp one, so the shift with warp order is almost identical on the two tables.
Their distances agree to $0.5~\Mpc$ ($0.04\,\sigma$) under the quadratic warp and $0.7~\Mpc$ ($0.08\,\sigma$) under the linear one, so the stronger curvature inferred on the \dsours table does not translate into a distance difference between the tables.

\begin{table*}
    \centering
    \caption{Per-galaxy disc parameters inferred under the quadratic warp on each of the two baseline spot tables, given as the posterior median and central $68\%$ interval.
    Warp rates and curvatures are evaluated at the pivot radius $r_\mathrm{ref}$ of~\cref{eq:warp_i}.
    \textit{Upper block}: the~\dsmcp table. \textit{Lower block}: the \dsours table, the same measurements under our own clipping (Section~\ref{sec:datasets}).
    As in~\cref{tab:disc_params}, the $\sigma^{(2)}$ rows are the NGC~5765b clump~2 error floors (Section~\ref{sec:disc_model}) and an em-dash marks a floor the model does not contain.}
    \label{tab:disc_params_qw}
    \setlength{\tabcolsep}{12pt}
    \renewcommand{\arraystretch}{1.3}
    \begin{tabular}{lccccc}
        \toprule
        Parameter                                                                   & CGCG~074-064                    & NGC~5765b                       & UGC~3789                        & NGC~6264                        & NGC~6323                        \\
        \midrule
        \multicolumn{6}{l}{\textit{\dsmcp table}} \\
        \midrule
        $\DA$ (\Mpc)                                                           & $85.7^{+8.0}_{-6.9}$      & $118.8^{+9.4}_{-8.7}$     & $50.7^{+4.6}_{-3.9}$      & $135^{+22}_{-18}$         & $154^{+58}_{-43}$         \\
        $\log(\MBH/\Msun)$                                                    & $7.374^{+0.039}_{-0.037}$ & $7.644^{+0.033}_{-0.033}$ & $7.069^{+0.038}_{-0.035}$ & $7.452^{+0.066}_{-0.062}$ & $7.16^{+0.14}_{-0.15}$    \\
        $i_0$ (\degunit)                                                           & $90.70^{+0.58}_{-0.57}$   & $84.65^{+0.30}_{-0.31}$   & $89.66^{+0.31}_{-0.28}$   & $90.41^{+0.45}_{-0.48}$   & $91.59^{+0.44}_{-0.58}$   \\
        $\mathrm{d}i/\mathrm{d}r|_{r_\mathrm{ref}}$ (\degmas)          & $6.1^{+7.3}_{-7.3}$       & $12.74^{+0.72}_{-0.80}$   & $9.10^{+1.03}_{-1.00}$    & $-2.1^{+6.3}_{-7.2}$      & $-12.0^{+9.2}_{-14.3}$    \\
        $\Omega_0$ (\degunit)                                                      & $101.27^{+0.39}_{-0.40}$  & $146.73^{+0.14}_{-0.12}$  & $221.50^{+0.11}_{-0.11}$  & $94.34^{+0.22}_{-0.22}$   & $189.72^{+0.14}_{-0.15}$  \\
        $\mathrm{d}\Omega/\mathrm{d}r|_{r_\mathrm{ref}}$ (\degmas)     & $20.9^{+5.3}_{-6.0}$      & $-4.15^{+0.53}_{-0.53}$   & $-0.23^{+1.00}_{-1.00}$   & $16.6^{+2.3}_{-2.3}$      & $17.6^{+2.2}_{-2.2}$      \\
        $\mathrm{d}^2i/\mathrm{d}r^2|_{r_\mathrm{ref}}$ (\degmassq)      & $4^{+31}_{-50}$           & $-1.2^{+3.8}_{-5.5}$      & $-11.5^{+3.6}_{-4.3}$     & $0^{+25}_{-26}$           & $-6^{+69}_{-86}$          \\
        $\mathrm{d}^2\Omega/\mathrm{d}r^2|_{r_\mathrm{ref}}$ (\degmassq) & $-31^{+12}_{-11}$         & $0.86^{+0.78}_{-1.46}$    & $-8.2^{+4.3}_{-4.3}$      & $-14^{+24}_{-24}$         & $-59^{+23}_{-27}$         \\
        $x_0$ (\muas)                                                       & $1.35^{+0.99}_{-1.06}$    & $-45.7^{+2.1}_{-2.1}$     & $-402.0^{+1.0}_{-1.0}$    & $4.8^{+1.2}_{-1.2}$       & $16.17^{+0.93}_{-0.93}$   \\
        $y_0$ (\muas)                                                       & $6.8^{+2.8}_{-2.8}$       & $-99.3^{+2.2}_{-2.3}$     & $-461.0^{+1.0}_{-1.1}$    & $7.9^{+1.7}_{-1.7}$       & $5.3^{+2.7}_{-2.7}$       \\
        $\Delta V_\mathrm{sys}$ ($\kmsec$)                                    & $-264.7^{+1.7}_{-2.0}$    & $-134.2^{+1.3}_{-1.2}$    & $14.73^{+0.82}_{-0.79}$   & $66.71^{+0.82}_{-0.79}$   & $190.0^{+1.9}_{-2.0}$     \\
        $\sigma_x$ (\muas)                                                  & $1.01^{+0.61}_{-0.36}$    & $4.8^{+2.6}_{-2.4}$       & $5.6^{+1.0}_{-1.1}$       & $1.32^{+0.94}_{-0.57}$    & $2.4^{+1.1}_{-1.0}$       \\
        $\sigma_y$ (\muas)                                                  & $14.0^{+2.5}_{-2.5}$      & $3.1^{+1.5}_{-1.4}$       & $5.1^{+1.4}_{-1.5}$       & $5.9^{+2.1}_{-2.0}$       & $3.8^{+2.5}_{-2.0}$       \\
        $\sigma_{v,\mathrm{sys}}$ ($\kmsec$)                                  & $1.73^{+0.86}_{-0.82}$    & $1.69^{+0.73}_{-0.74}$    & $1.74^{+0.91}_{-0.83}$    & $1.64^{+0.88}_{-0.80}$    & $2.01^{+0.98}_{-0.92}$    \\
        $\sigma_{v,\mathrm{hv}}$ ($\kmsec$)                                   & $2.13^{+0.94}_{-0.89}$    & $1.98^{+0.85}_{-0.85}$    & $1.78^{+0.92}_{-0.85}$    & $1.26^{+0.68}_{-0.62}$    & $1.76^{+0.91}_{-0.85}$    \\
        $\sigma_a$ ($\kmsecyr$)                                               & $0.24^{+0.11}_{-0.11}$    & $0.050^{+0.022}_{-0.018}$ & $0.286^{+0.073}_{-0.074}$ & $0.084^{+0.059}_{-0.042}$ & $0.169^{+0.092}_{-0.084}$ \\
        \midrule
        \multicolumn{6}{l}{\textit{\dsours table}} \\
        \midrule
        $\DA$ (\Mpc) & $86.0^{+8.0}_{-7.2}$ & $119.3^{+8.5}_{-7.6}$ & $50.8^{+4.3}_{-3.7}$ & $141^{+24}_{-19}$ & $125^{+46}_{-29}$ \\
        $\log(\MBH/\Msun)$ & $7.376^{+0.039}_{-0.039}$ & $7.648^{+0.030}_{-0.029}$ & $7.069^{+0.035}_{-0.033}$ & $7.471^{+0.070}_{-0.064}$ & $7.07^{+0.14}_{-0.12}$ \\
        $i_0$ (\degunit) & $90.49^{+0.40}_{-0.41}$ & $85.04^{+0.18}_{-0.19}$ & $89.67^{+0.29}_{-0.26}$ & $90.36^{+0.47}_{-0.48}$ & $91.69^{+0.33}_{-0.38}$ \\
        $\mathrm{d}i/\mathrm{d}r|_{r_\mathrm{ref}}$ (\degmas) & $-0.0^{+5.9}_{-5.4}$ & $11.13^{+0.72}_{-0.82}$ & $8.79^{+0.94}_{-0.91}$ & $-1.4^{+6.2}_{-6.4}$ & $-6.5^{+7.2}_{-5.5}$ \\
        $\Omega_0$ (\degunit) & $101.78^{+0.32}_{-0.32}$ & $146.90^{+0.11}_{-0.11}$ & $221.459^{+0.089}_{-0.087}$ & $94.34^{+0.23}_{-0.22}$ & $189.66^{+0.12}_{-0.12}$ \\
        $\mathrm{d}\Omega/\mathrm{d}r|_{r_\mathrm{ref}}$ (\degmas) & $19.4^{+5.1}_{-5.5}$ & $-4.42^{+0.51}_{-0.52}$ & $-0.91^{+0.73}_{-0.73}$ & $16.5^{+2.2}_{-2.1}$ & $16.1^{+2.1}_{-2.1}$ \\
        $\mathrm{d}^2i/\mathrm{d}r^2|_{r_\mathrm{ref}}$ (\degmassq) & $33^{+16}_{-64}$ & $-6.54^{+0.89}_{-0.92}$ & $-10.6^{+3.3}_{-3.9}$ & $5^{+23}_{-23}$ & $8^{+36}_{-45}$ \\
        $\mathrm{d}^2\Omega/\mathrm{d}r^2|_{r_\mathrm{ref}}$ (\degmassq) & $-31^{+11}_{-10}$ & $-0.40^{+0.57}_{-0.55}$ & $-6.9^{+1.5}_{-1.5}$ & $-17^{+23}_{-23}$ & $-28^{+15}_{-16}$ \\
        $x_0$ (\muas) & $1.71^{+0.90}_{-0.96}$ & $-47.0^{+2.0}_{-2.0}$ & $-401.78^{+0.91}_{-0.92}$ & $4.4^{+1.2}_{-1.3}$ & $15.73^{+0.90}_{-0.91}$ \\
        $y_0$ (\muas) & $5.6^{+2.3}_{-2.3}$ & $-98.3^{+2.2}_{-2.3}$ & $-461.27^{+0.99}_{-1.00}$ & $7.4^{+1.7}_{-1.7}$ & $6.6^{+2.5}_{-2.5}$ \\
        $\Delta V_\mathrm{sys}$ ($\kmsec$) & $-264.0^{+1.6}_{-1.8}$ & $-132.6^{+1.1}_{-1.1}$ & $14.77^{+0.78}_{-0.76}$ & $66.92^{+0.91}_{-0.89}$ & $191.2^{+1.6}_{-1.7}$ \\
        $\sigma_x$ (\muas) & $1.03^{+0.63}_{-0.38}$ & $6.9^{+2.0}_{-2.2}$ & $4.39^{+0.97}_{-1.02}$ & $1.56^{+1.22}_{-0.74}$ & $1.60^{+0.88}_{-0.71}$ \\
        $\sigma_y$ (\muas) & $4.6^{+2.8}_{-2.5}$ & $3.4^{+1.7}_{-1.6}$ & $4.7^{+1.3}_{-1.4}$ & $6.1^{+2.0}_{-1.9}$ & $3.5^{+2.2}_{-1.8}$ \\
        $\sigma_{v,\mathrm{sys}}$ ($\kmsec$) & $1.80^{+0.88}_{-0.87}$ & $1.83^{+0.86}_{-0.86}$ & $1.82^{+0.89}_{-0.87}$ & $1.67^{+0.89}_{-0.83}$ & $2.03^{+0.98}_{-0.95}$ \\
        $\sigma_{v,\mathrm{hv}}$ ($\kmsec$) & $2.12^{+0.93}_{-0.92}$ & $3.01^{+0.81}_{-0.89}$ & $1.90^{+0.92}_{-0.87}$ & $2.07^{+0.69}_{-0.72}$ & $1.78^{+0.94}_{-0.87}$ \\
        $\sigma_a$ ($\kmsecyr$) & $0.27^{+0.12}_{-0.12}$ & $0.085^{+0.023}_{-0.019}$ & $0.290^{+0.070}_{-0.070}$ & $0.124^{+0.056}_{-0.049}$ & $0.178^{+0.093}_{-0.086}$ \\
        $\sigma_x^{(2)}$ (\muas) & --- & $11.6^{+2.2}_{-2.9}$ & --- & --- & --- \\
        $\sigma_y^{(2)}$ (\muas) & --- & $5.0^{+2.5}_{-2.1}$ & --- & --- & --- \\
        $\sigma_{v,\mathrm{sys}}^{(2)}$ ($\kmsec$) & --- & $2.04^{+0.89}_{-0.90}$ & --- & --- & --- \\
        $\sigma_a^{(2)}$ ($\kmsecyr$) & --- & $0.187^{+0.083}_{-0.074}$ & --- & --- & --- \\
        \bottomrule
    \end{tabular}
\end{table*}

\section{Example per-galaxy posterior}\label{app:corner}

\begin{figure*}
    \centering
    \includegraphics[width=0.92\textwidth]{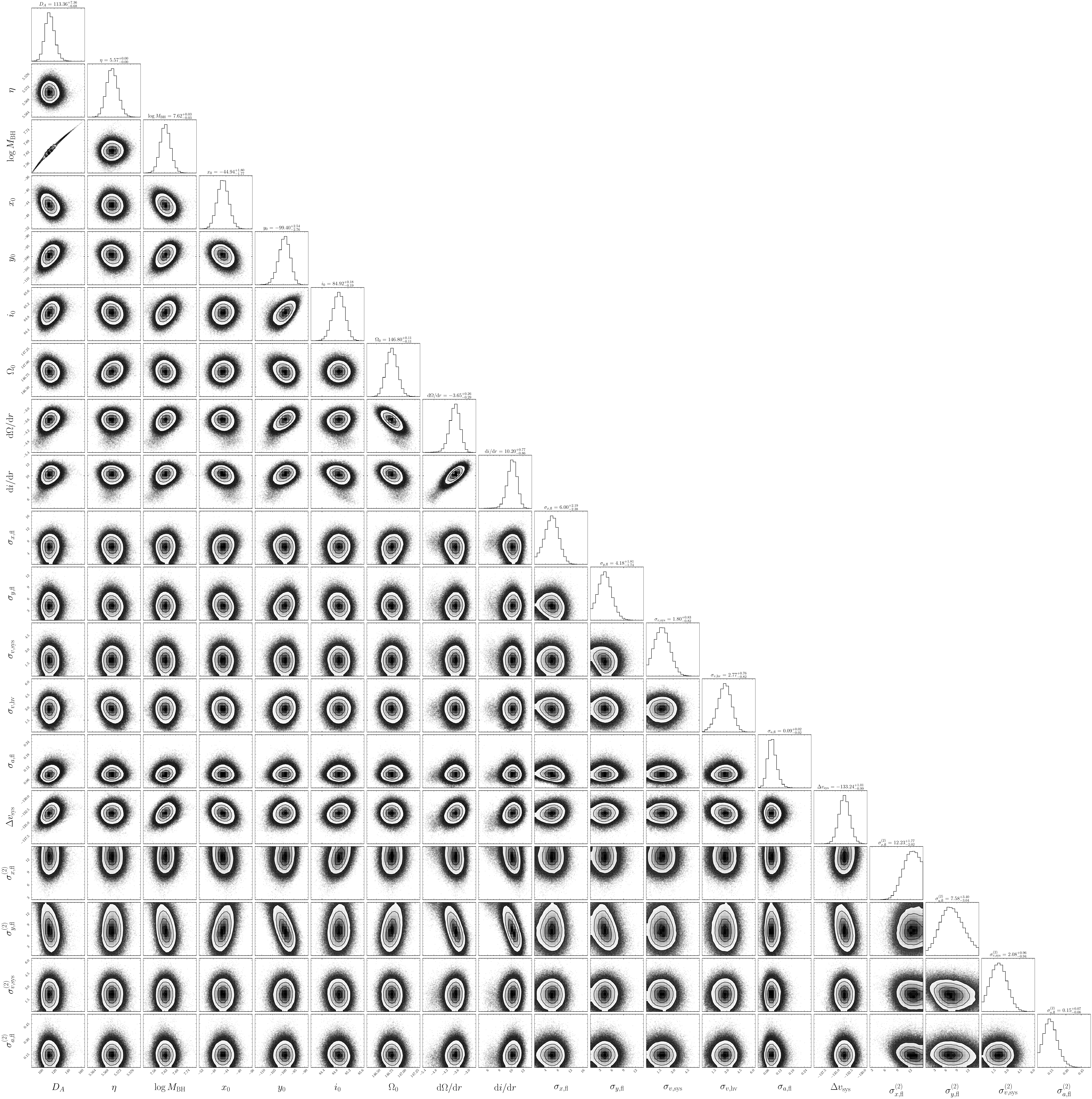}
    \caption{Posterior on the global disc parameters of NGC~5765b for the baseline linear-warp model on the \dsours table, typical of the five galaxies.
    Contours are the $68\%$ and $95\%$ credible regions.
    The four $\sigma^{(2)}$ parameters are the clump~2 error floors, which the model contains only on the tables the MCP cut is not applied to.
    All global disc parameters have $\hat{R} < 1.01$, with ${\approx}6.2\times10^{3}$ effective samples on $\DA$.
    }
    \label{fig:corner_ngc5765b}
\end{figure*}

\Cref{fig:corner_ngc5765b} shows a strong correlation between the angular-diameter distance of NGC~5765b and its black hole mass.
We do not sample this pair, but rather $\eta$ and $\DA$, which are almost uncorrelated by construction.

\section{NGC~4258 disc-model variants}\label{app:ngc4258}

\begin{figure*}
    \centering
    \includegraphics[width=0.92\textwidth]{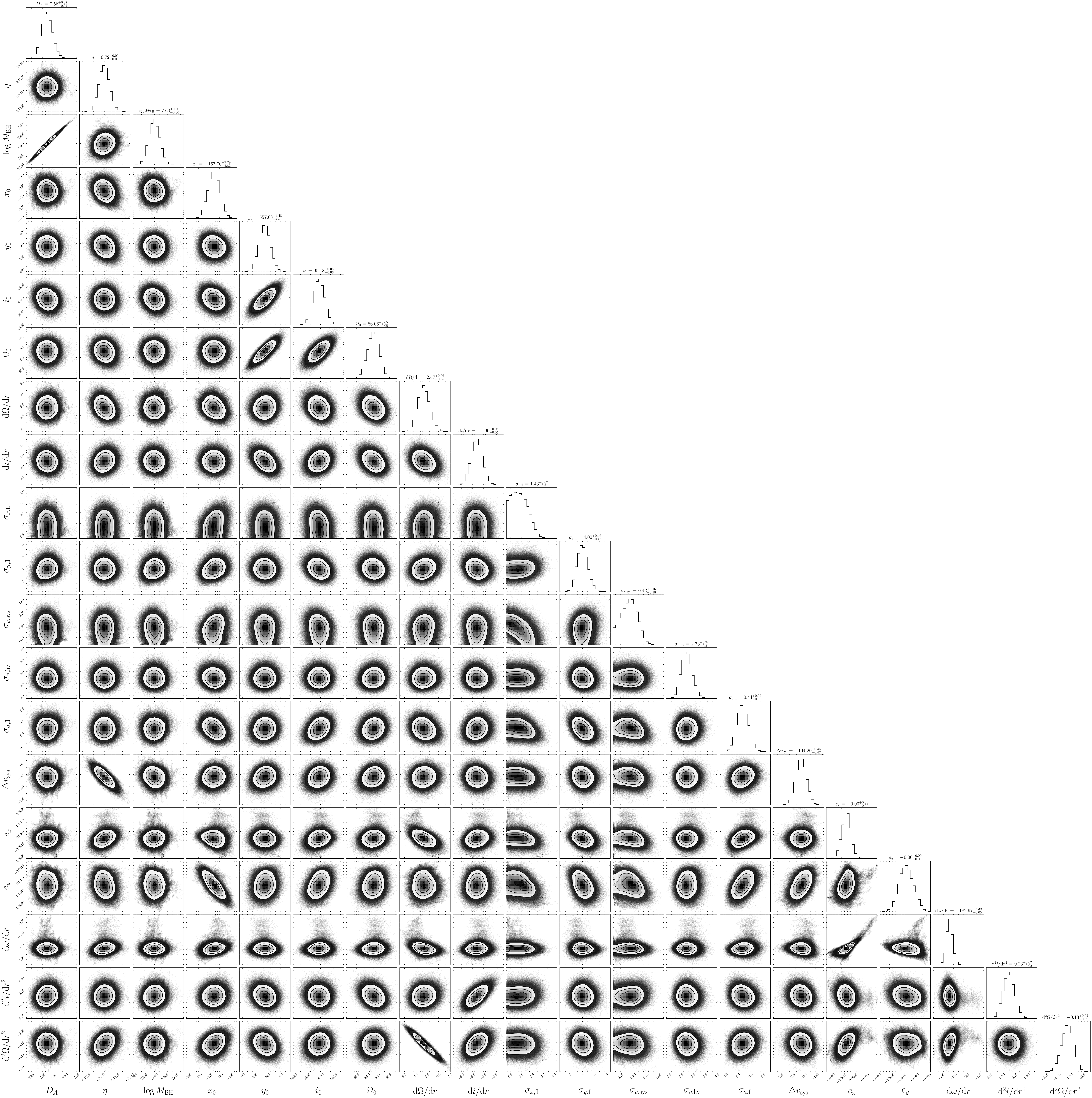}
    \caption{Posterior on the global disc parameters of NGC~4258 under the eccentric quadratic-warp model, inferred from the full $358$-spot table with no outlier rejection.
    Contours are the $68\%$ and $95\%$ credible regions.
    The eccentricity enters through the Cartesian components $e_x$ and $e_y$, with $e = (e_x^2 + e_y^2)^{1/2} = 0.0041 \pm 0.0009$, and $\mathrm{d}\omega/\mathrm{d}r$ is the periapsis warp rate.}
    \label{fig:corner_ngc4258_ecc}
\end{figure*}

\Cref{tab:ngc4258_variants} reports the NGC~4258 disc posteriors under the quadratic warp and its eccentric extension, both on the full $358$-spot table with no outlier rejection.
The two models differ by $0.046~\Mpc$ in $\DA$ and both lie within $0.51\,\sigma$ of the~\citet{Reid2019} distance.
The disc geometry is likewise stable: between the two, the inclination and position angle move by less than $0.04$ and $0.02~\degunit$, respectively, and the curvature coefficients by less than $1\,\sigma$.
\Cref{fig:corner_ngc4258_ecc} shows the corresponding posterior for the eccentric model.

\begin{table*}
    \centering
    \caption{Disc parameters of NGC~4258 inferred under the quadratic-warp model and its eccentric extension, given as the posterior median and central $68\%$ interval.
    Warp rates and curvatures are evaluated at the pivot radius $r_\mathrm{ref}$ of~\cref{eq:warp_i} and the periapsis warp rate at $r_\mathrm{ref}^{\omega}$, both $5.1~\mathrm{mas}$ for NGC~4258, approximately the median orbital radius of its maser spots.
    The eccentricity $e = (e_x^2 + e_y^2)^{1/2}$ and the argument of periapsis $\omega_0 = \mathrm{atan2}(e_y, e_x)$ are derived from the sampled components $e_x$ and $e_y$ (\cref{tab:priors}).}
    \label{tab:ngc4258_variants}
    \setlength{\tabcolsep}{10pt}
    \renewcommand{\arraystretch}{1.3}
    \begin{tabular}{lcc}
        \toprule
        Parameter & Quadratic & Quadratic \\
                  &           & $+$ eccentricity \\
        \midrule
        Spots & $358$ & $358$ \\
        $\DA$ (\Mpc) & $7.512^{+0.061}_{-0.055}$ & $7.558^{+0.074}_{-0.072}$ \\
        $\log(\MBH/\Msun)$ & $7.597^{+0.003}_{-0.003}$ & $7.600^{+0.004}_{-0.004}$ \\
        $i_0$ (\degunit) & $95.82^{+0.05}_{-0.05}$ & $95.78^{+0.06}_{-0.06}$ \\
        $\mathrm{d}i/\mathrm{d}r|_{r_\mathrm{ref}}$ (\degmas) & $-1.98^{+0.05}_{-0.05}$ & $-1.96^{+0.05}_{-0.05}$ \\
        $\mathrm{d}^2i/\mathrm{d}r^2|_{r_\mathrm{ref}}$ (\degmassq) & $0.222^{+0.020}_{-0.019}$ & $0.227^{+0.023}_{-0.022}$ \\
        $\Omega_0$ (\degunit) & $86.07^{+0.05}_{-0.05}$ & $86.06^{+0.05}_{-0.05}$ \\
        $\mathrm{d}\Omega/\mathrm{d}r|_{r_\mathrm{ref}}$ (\degmas) & $2.50^{+0.04}_{-0.04}$ & $2.47^{+0.06}_{-0.05}$ \\
        $\mathrm{d}^2\Omega/\mathrm{d}r^2|_{r_\mathrm{ref}}$ (\degmassq) & $-0.141^{+0.015}_{-0.014}$ & $-0.125^{+0.019}_{-0.020}$ \\
        $e$ & --- & $0.0041^{+0.0009}_{-0.0010}$ \\
        $\omega_0$ (\degunit) & --- & $257^{+8}_{-7}$ \\
        $\mathrm{d}\omega/\mathrm{d}r|_{r_\mathrm{ref}^{\omega}}$ (\degmas) & --- & $-183^{+6}_{-6}$ \\
        $x_0$ (\muas) & $-177.7^{+1.6}_{-1.7}$ & $-167.7^{+2.8}_{-2.8}$ \\
        $y_0$ (\muas) & $560^{+4}_{-4}$ & $558^{+4}_{-5}$ \\
        $\Delta V_\mathrm{sys}$ ($\kmsec$) & $-193.2^{+0.4}_{-0.4}$ & $-194.2^{+0.5}_{-0.5}$ \\
        $\sigma_x$ (\muas) & $0.93^{+0.40}_{-0.30}$ & $1.43^{+0.67}_{-0.61}$ \\
        $\sigma_y$ (\muas) & $3.3^{+0.4}_{-0.3}$ & $4.0^{+0.5}_{-0.4}$ \\
        $\sigma_{v,\mathrm{sys}}$ ($\kmsec$) & $0.27^{+0.11}_{-0.10}$ & $0.42^{+0.16}_{-0.18}$ \\
        $\sigma_{v,\mathrm{hv}}$ ($\kmsec$) & $3.02^{+0.24}_{-0.22}$ & $2.73^{+0.24}_{-0.21}$ \\
        $\sigma_a$ ($\kmsecyr$) & $0.47^{+0.04}_{-0.04}$ & $0.44^{+0.05}_{-0.05}$ \\
        \bottomrule
    \end{tabular}
\end{table*}

\section{Joint posteriors of the population runs}\label{app:population_corners}

\begin{figure*}
    \centering
    \includegraphics[width=0.92\textwidth]{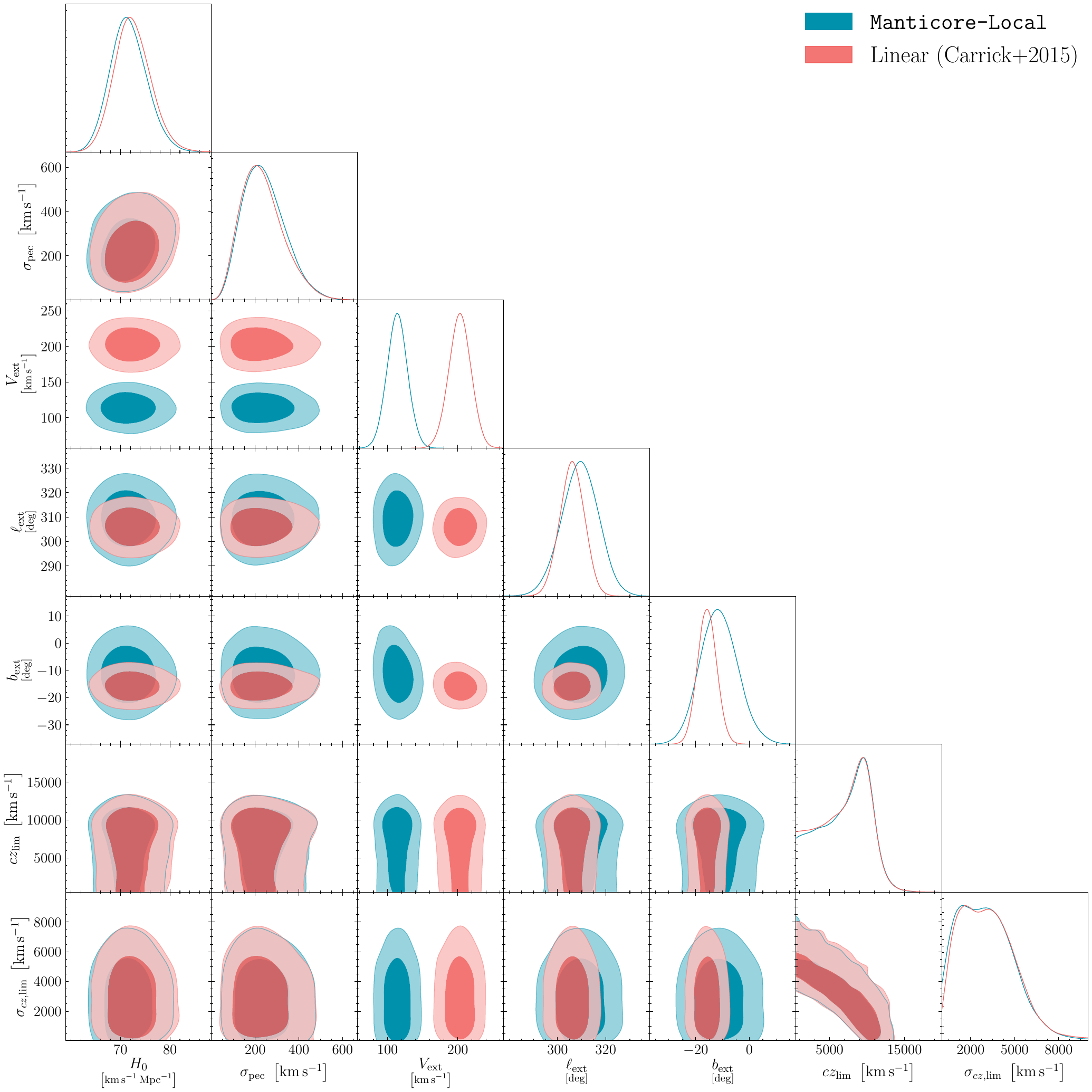}
    \caption{Joint posterior over the shared population parameters for the redshift-selection runs on the \dsours baseline spot table (Section~\ref{sec:datasets}), overlaying the \Manticore\ (blue) and linear (red) reconstructions, both with a linear warp.
        Contours are the $68\%$ and $95\%$ credible regions.
        The two reconstructions give consistent $H_0$, while $\sigma_\mathrm{pec}$ and $\Vext$ recover their priors.}
    \label{fig:corner_h0_redshift}
\end{figure*}

\begin{figure*}
    \centering
    \includegraphics[width=0.92\textwidth]{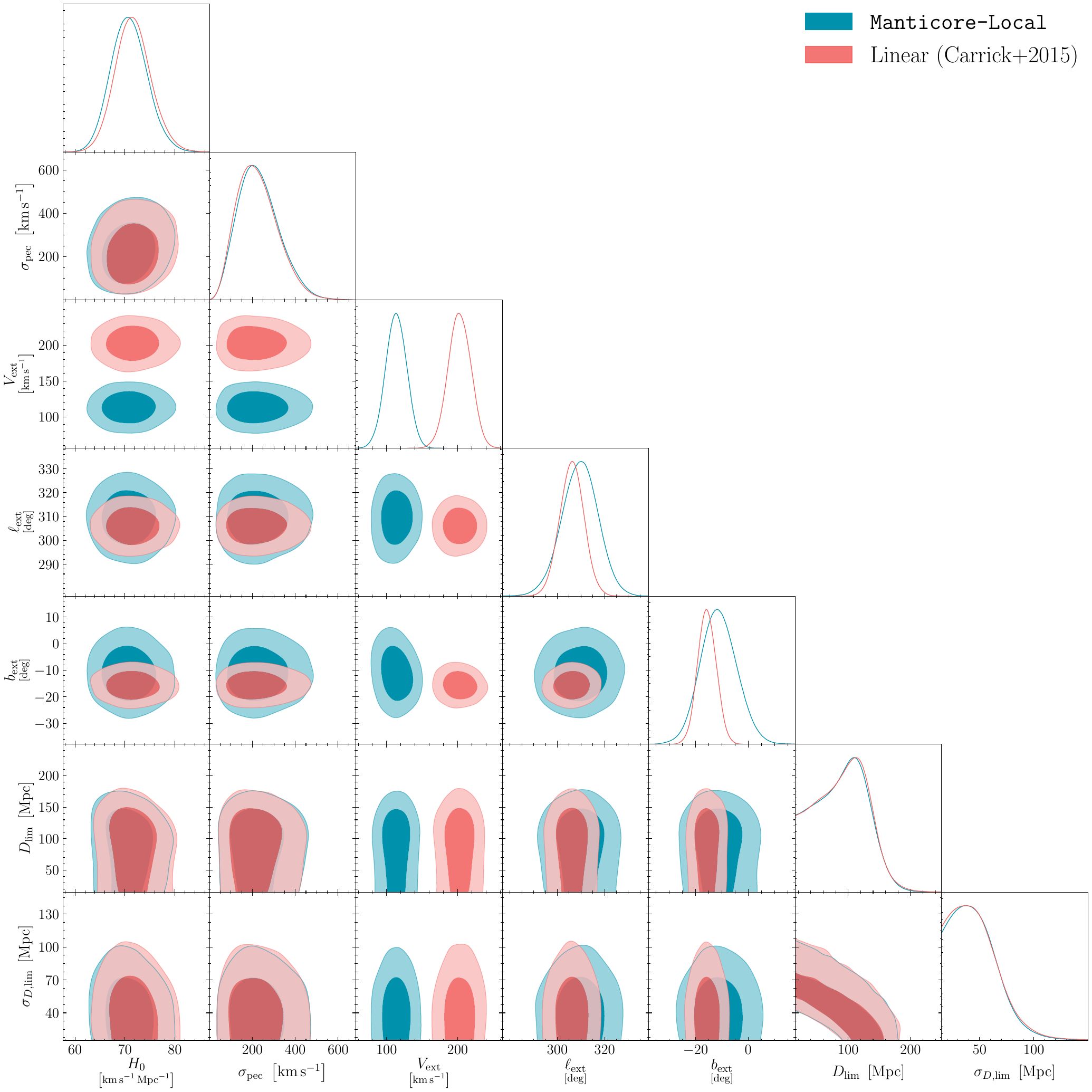}
    \caption{As in~\cref{fig:corner_h0_redshift}, but for the distance-selection runs, with the redshift threshold replaced by the distance threshold $(D_\mathrm{lim}, D_\mathrm{width})$.}
    \label{fig:corner_h0_distance}
\end{figure*}

\Cref{fig:corner_h0_redshift,fig:corner_h0_distance} show the joint posteriors over the shared population parameters for the two selection runs that include a reconstruction, overlaying the \Manticore\ and linear~\citepalias{Carrick2015} reconstructions, both with a linear warp, on the \dsours baseline table.
We sample $\Vext$ in Cartesian components and show it here as its magnitude $V_\mathrm{ext}$ and Galactic direction.
The two reconstructions yield a near-identical, prior-dominated $\sigma_\mathrm{pec}$ and $H_0$ values consistent within their uncertainties, with the linear field higher by ${\sim}0.8~\kmsecMpc$.
In both runs the five galaxies do not constrain the residual bulk flow $\Vext$, its posterior simply recovering the informative Gaussian prior.
The selection thresholds are only loosely constrained, the posterior on the redshift limit $cz_\mathrm{lim}$ peaking near the redshift of the most distant galaxy with a broad $cz_\mathrm{width}$, and the distance threshold $(D_\mathrm{lim}, D_\mathrm{width})$ similarly wide.

\end{document}